\documentclass[sn-mathphys-num]{sn-jnl}

\usepackage{graphicx}%
\usepackage{multirow}%
\usepackage{amsmath,amssymb,amsfonts}%
\usepackage{amsthm}%
\usepackage{mathrsfs}%
\usepackage[title]{appendix}%
\usepackage{xcolor}%
\usepackage{textcomp}%
\usepackage{manyfoot}%
\usepackage{booktabs}%
\usepackage{algorithm}%
\usepackage{algorithmicx}%
\usepackage{algpseudocode}%
\usepackage{listings}%
\usepackage{lineno}
\usepackage{xspace}

\usepackage{epsfig}
\usepackage{dcolumn}
\usepackage{amssymb}   
\usepackage{amsmath}
\usepackage{mathrsfs}
\usepackage{bm}
\usepackage{ulem}
\usepackage{graphicx}
\usepackage{subfigure}
\usepackage{pstricks}
\usepackage{pst-node}
\usepackage{rotating}
\usepackage{times}
\usepackage{soul}
\usepackage{float}
\usepackage[english]{babel}
\addto{\captionsenglish}{%
  \renewcommand{\figurename}{FIG.}
  \renewcommand{\tablename}{TABLE}
}
\include{bes3physics.sty}

\begin{document}

\title[Article Title]{Studies of the tracking and identification efficiencies of electrons and positrons at BESIII}

\author*[1,2]{\fnm{Xinyu} \sur{Chai}}\email{chaixy@stu.pku.edu.cn}

\author[1,2]{\fnm{Mengzhen} \sur{Wang}}

\author[3]{\fnm{Xiaobin} \sur{Ji}}

\author[3]{\fnm{Shengsen} \sur{Sun}}

\author*[1,2]{\fnm{Dayong} \sur{Wang}}\email{dayong.wang@pku.edu.cn}

\affil*[1]{\orgdiv{State Key Laboratory of Nuclear Physics and Technology}, \orgname{Peking University}, \orgaddress{\street{}\city{Beijing}, \postcode{100871}, \state{} \country{China}}}

\affil*[2]{\orgdiv{School of Physics}, \orgname{Peking University}, \orgaddress{\street{}\city{Beijing},  \postcode{100871}, \state{}\country{China}}}

\affil[3]{\orgdiv{}\orgname{Institute of High Energy Physics}, \orgaddress{\street{}\city{Beijing},  \postcode{100049}, \state{}\country{China}}}



\abstract{The efficiencies for electron and positron tracking and identification in the BESIII experiment are investigated with the radiative Bhabha process $e^+e^-\ar e^+e^-\gamma$ from the data samples collected at the center-of-mass energies of 3.08 GeV and 3.097 GeV. The relative differences between data and MC associated with tracking and identification efficiencies of electrons and positrons, as well as the corresponding correction factors 
are determined. It turns out the relative differences of tracking efficiency and particle identification efficiency after correction are mostly less than 0.5$\%$ for transverse momenta $p_T>0.4$~GeV and for the entire momentum region, respectively. }




\maketitle

\section{Introduction}\label{sec1}
The BESIII experiment is a $\tau$-charm factory with a peak luminosity of $1.1\times10^{33}$ cm$^{-2}$s$^{-1}$ running at the center-of-mass energy ($\sqrt s$) in the range between 1.84 and 4.95 GeV~\cite{Asner:2008nq, BESIII:2020nme}. 
As large data sets have been accumulated, statistical uncertainties decrease significantly while systematic uncertainties become increasingly important, dominant in some cases, for precision measurements.
The detection and identification of electrons plays a key role in many experimental studies, such as semi-leptonic decays, purely leptonic decays, and electro-magnetic Dalitz decays.
Therefore, it is essential to systematically investigate electron tracking and identification for robust data analysis and comprehensive understanding of detector performance.   Throughout this paper, charge conjugation is always implied unless explicitly noted.

The key tracking sub-detector of the BESIII detector~\cite{Ablikim:2009aa} is a small-celled, helium-based multi-layer drift chamber (MDC) with 43 layers of wires, which has a geometrical acceptance of 93$\%$ of $4\pi$ and provides momentum and ionization energy loss (d$E$/d$x$) measurements for charged particles. The other sub-detectors include a plastic scintillator time-of-flight (TOF) system, a CsI (Tl) electromagnetic calorimeter (EMC) and a muon counter made of Resistive Plate Chambers in the iron flux return yoke of the superconducting magnet. 
The TOF system provides time-of-flight measurements of charged particles from the $\ee$ collision point to the impact point in the TOF detector, and the EMC measures the deposited energy of the charged particle.  Both measurements are useful information for particle identification (PID).
The charged-particle momentum resolution at $1~{\rm GeV}/c$ is $0.5\%$, and the ${\rm d}E/{\rm d}x$ resolution is $6\%$ for electrons from Bhabha scattering. The EMC measures photon energies with a resolution of $2.5\%$ ($5\%$) at $1$~GeV in the barrel (end cap) region. The time resolution in the plastic scintillator TOF barrel region is 68~ps, while that in the end cap region was 110~ps. The end cap TOF system was upgraded in 2015 using multigap resistive plate chamber technology, providing a time resolution of
60~ps, which benefits all the data used in this study~\cite{etof}.

In experiments, the systematic uncertainties mostly arise from discrepancies between data and Monte Carlo (MC) simulation.
In this study, the optimization of the event selection, the investigation of possible backgrounds, and various validations  are performed using MC simulated samples. These samples are produced with a {\sc
geant4}-based~\cite{geant4} software package, which
includes the geometric description of the BESIII detector and its response. The inclusive MC sample includes both the production of the $J/\psi$
resonance and the continuum processes incorporated in {\sc
kkmc}~\cite{ref:kkmc}.
For the tracking efficiency study, MC simulated events of the radiative Bhabha process
$\ee\ar\ee\gamma$ are generated at $\sqrt s=3.08 \gev$ using the BABAYAGA$@$NLO event generator~\cite{BabayagaNLO_1, BabayagaNLO_2}. 
A mixed sample of radiative Bhabha process $\ee\ar\ee\gamma$, generated using the BABAYAGA3.5 event generator \cite{Babayaga3_1, Babayaga3_2}, and the $\jpsi\ar\ee$ process with photon from final state radiation generated using the PHOTOS VLL generator, is used to study the PID efficiency. 
Final state radiation
from charged final state particles from $J/\psi$ decays is  incorporated using {\sc
photos}~\cite{photos2}.

In this paper, the systematic uncertainties associated with the electron tracking and PID efficiencies at BESIII are studied. 
High statistics and pure electron sample is selected using the radiative Bhabha scattering process, providing a broad range of momentum and angular distributions. 
The tracking efficiency study is based on a data sample with an integrated luminosity of (136.22 $\pm$ 0.09)~pb$^{-1}$ collected at $\sqrt s=3.08 \gev$~\cite{BESIII:2021cxx}; and the PID efficiency study is based on a data sample with 
an integrated luminosity of (2568.07 $\pm$ 0.40) pb$^{-1}$ 
collected at $\sqrt s=3.097 \gev$~\cite{BESIII:2021cxx}.

\section{Definitions of efficiency, correction factor, and uncertainty}\label{sec2}
The tag-and-probe method is used to determine the tracking and PID efficiencies of both data and MC samples. 
The tagged positron is required to satisfy the positron selection criteria described in Secs.~\ref{sec::trk} and \ref{sec::pid}. In the tracking efficiency study, the probed electron is treated as a missing track which can only be inferred from information on the recoil side. In the PID efficiency study, the probed electron is required to satisfy further PID selection criteria. Subsequently, within the selected control samples, the number of successfully  reconstructed and identified probed positron can be quantified and used in the efficiency calculation.
The electron tracking or identification efficiency ($\epsilon$) is defined as 
\begin{equation}
    \epsilon = \frac{n}{N},
    \label{eq_eff}
\end{equation}
where $N$ is the number of events determined just by the tagged positron; $n$ is the number of events where the probed electron is found or identified successfully. The statistical uncertainty of the efficiency is determined by
\begin{equation}
    \sigma=\sqrt{\frac{\epsilon(1-\epsilon)}{N}}.
    \label{eq_eff_sys}
\end{equation}
The relative difference of tracking or PID efficiencies between MC ($\epsilon_{\rm MC}$) and data ($\epsilon_{\rm data}$) is defined as
\begin{equation}
    \Delta_\epsilon=1-\frac{\epsilon_{\rm data}}{\epsilon_{\rm MC}}.
    \label{eq_correction}
\end{equation}
The correction factor ($\alpha$) for tracking or PID efficiency, to be applied to MC simulations, is defined as 
\begin{equation}
  \alpha = 1-\Delta_\epsilon = \frac{\epsilon_{\rm data}}{\epsilon_{\rm MC}},  
\end{equation}
which can be used to reweight the corresponding signal MC sample. The uncertainties of the correction factor and relative difference are the same; therefore, we use $\sigma_{\alpha}$ as a unified representation for both quantities. Due to the independence of $\epsilon_{\rm MC}$ and $\epsilon_{\rm data}$, $\sigma_{\alpha}$ is defined as
\begin{equation}
    \sigma_{\alpha}=\frac{\epsilon_{\rm data}}{\epsilon_{\rm MC}}\sqrt{\frac{\sigma^2_{\rm MC}}{\epsilon^2_{\rm MC}}+\frac{\sigma^2_{\rm data}}{\epsilon^2_{\rm data}}}.
    \label{eq_correction_sys}
\end{equation}
Because the same event selection criteria applied when obtaining $N$ and $n$, the systematic uncertainty caused by event selection criteria cancel out.

\section{Tracking efficiency study}

\subsection{Control sample selection}\label{sec::trk}
The candidate events of $\ee\ar\ee\gamma$ are required to have at least one good charged track. The good charged track must be reconstructed in the MDC within its angular coverage, $|\cos\theta| < 0.93$, where $\theta$ is the polar angle with respect to the $z$-axis,
which is the symmetry axis of the MDC. For electron candidates, their 
distance of closest approach to the interaction point  must be less than 10\,cm along the $z$-axis,
and less than 1\,cm in the transverse plane.
A good charged track with momentum greater than 1.46 GeV, which corresponds to $95\%$ of the beam momentum, is considered as the tagged track. For the tagged track, it is identified as the electron if the ratio of energy deposited in the EMC over its momentum ($E/p$) is within the range from 0.8 to 1.2. Only one tagged electron is required. 

\begin{figure*}[!htbp]
\centering
\subfigure{
\centering
\includegraphics[width=0.4\textwidth]{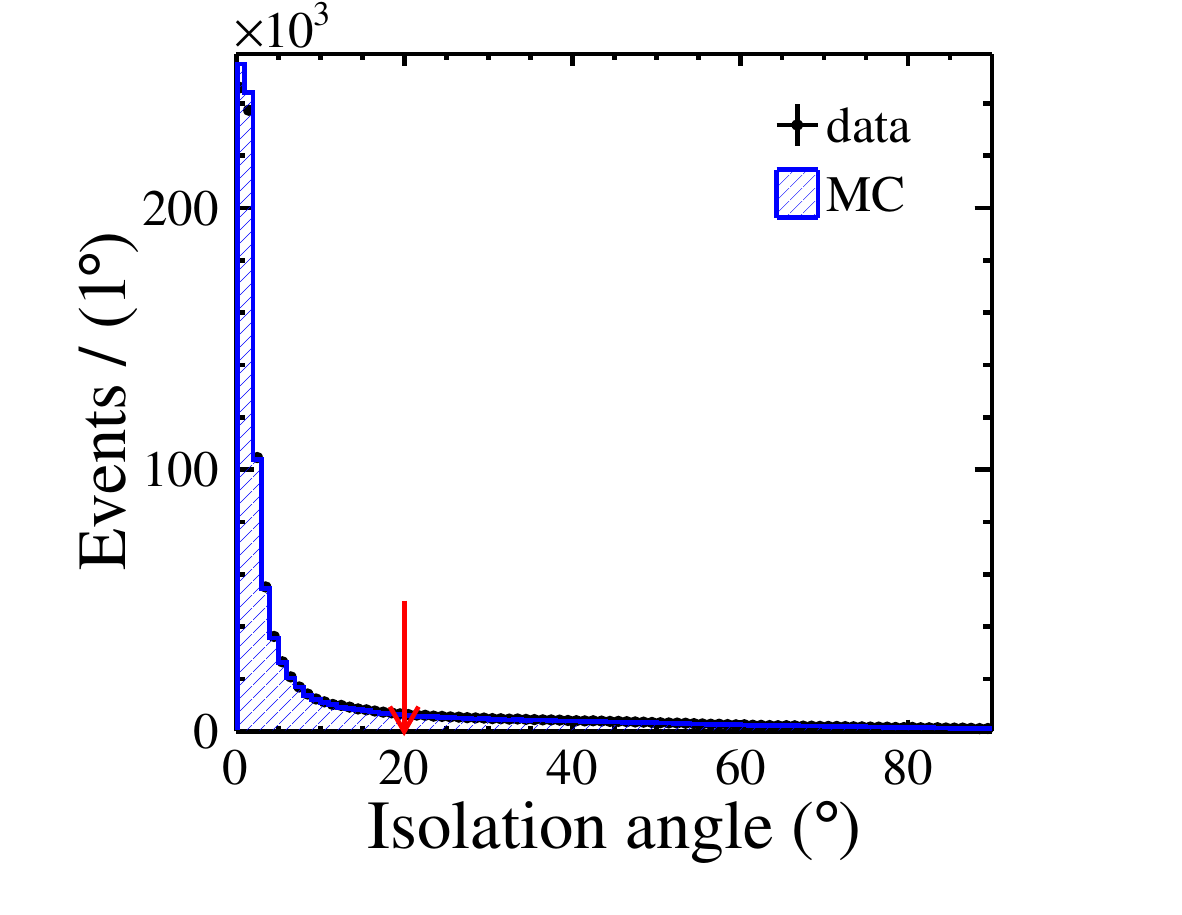}
}%
\subfigure{
\centering
\includegraphics[width=0.4\textwidth]{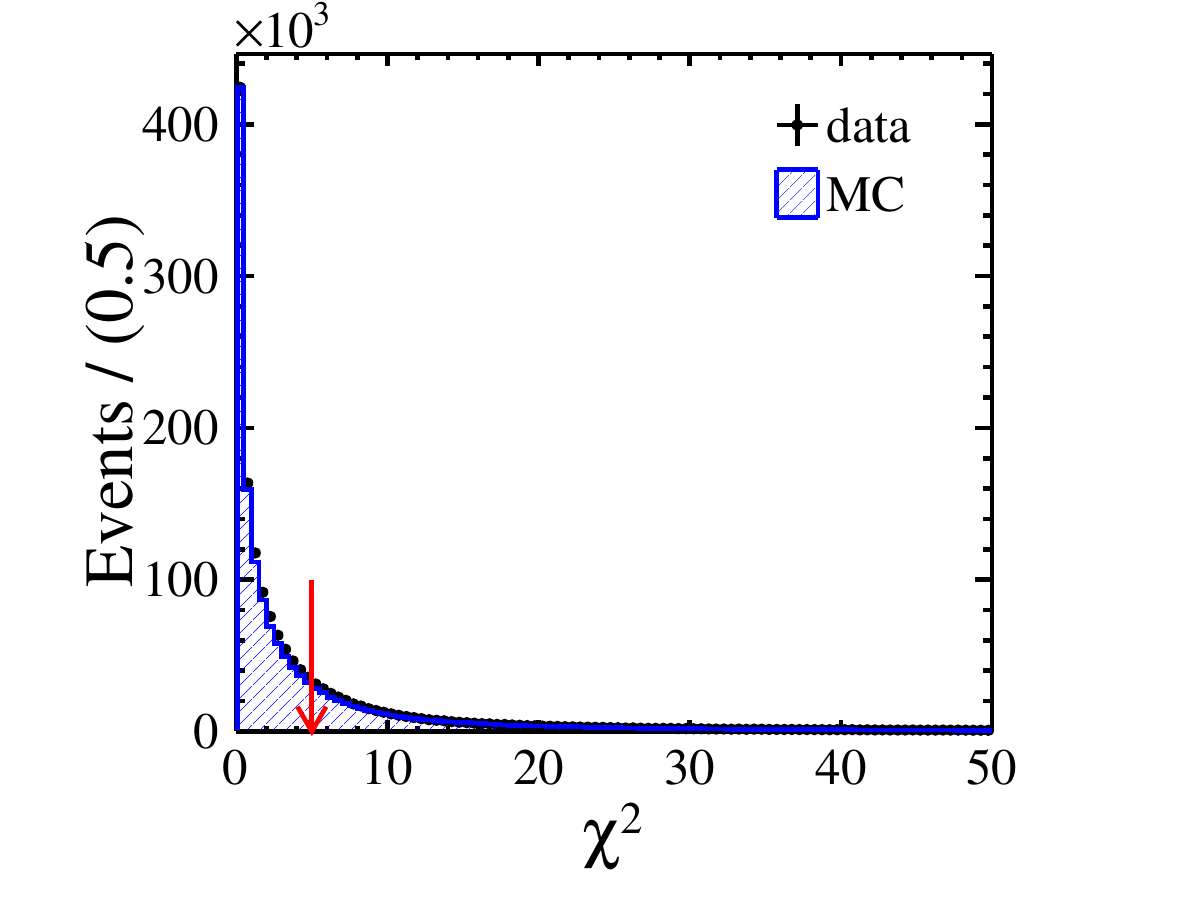}
}%
\caption{\small Distributions of the isolation angle (left) and $\chi^2$ (right) for selected data and MC samples in the tracking efficiency study.}
\label{evt_tracking}
\end{figure*}

The candidate events must also have at least one good photon. Good photon candidates are reconstructed by using isolated showers in the EMC.  
The deposited energy of each shower must be more than 25~MeV in the barrel regions ($|\cos \theta|< 0.80$) and more than 50~MeV in the end cap region ($0.86 <|\cos \theta|< 0.92$). 
Since the tagged track has higher momentum, most radiative photons originate from other tracks and thus have lower momenta. 
To select radiative Bhabha events, the opening angle between the photon and the probed electron is required to be less than $20^\circ$, as shown in Fig.~\ref{evt_tracking}.
To suppress electronic noise and showers unrelated to the event, the difference between the EMC time and the event start time is required to be within [0, 700]\,ns. 

\begin{figure*}[!htbp]
\centering
\subfigure{
\centering
\includegraphics[width=0.3\textwidth]{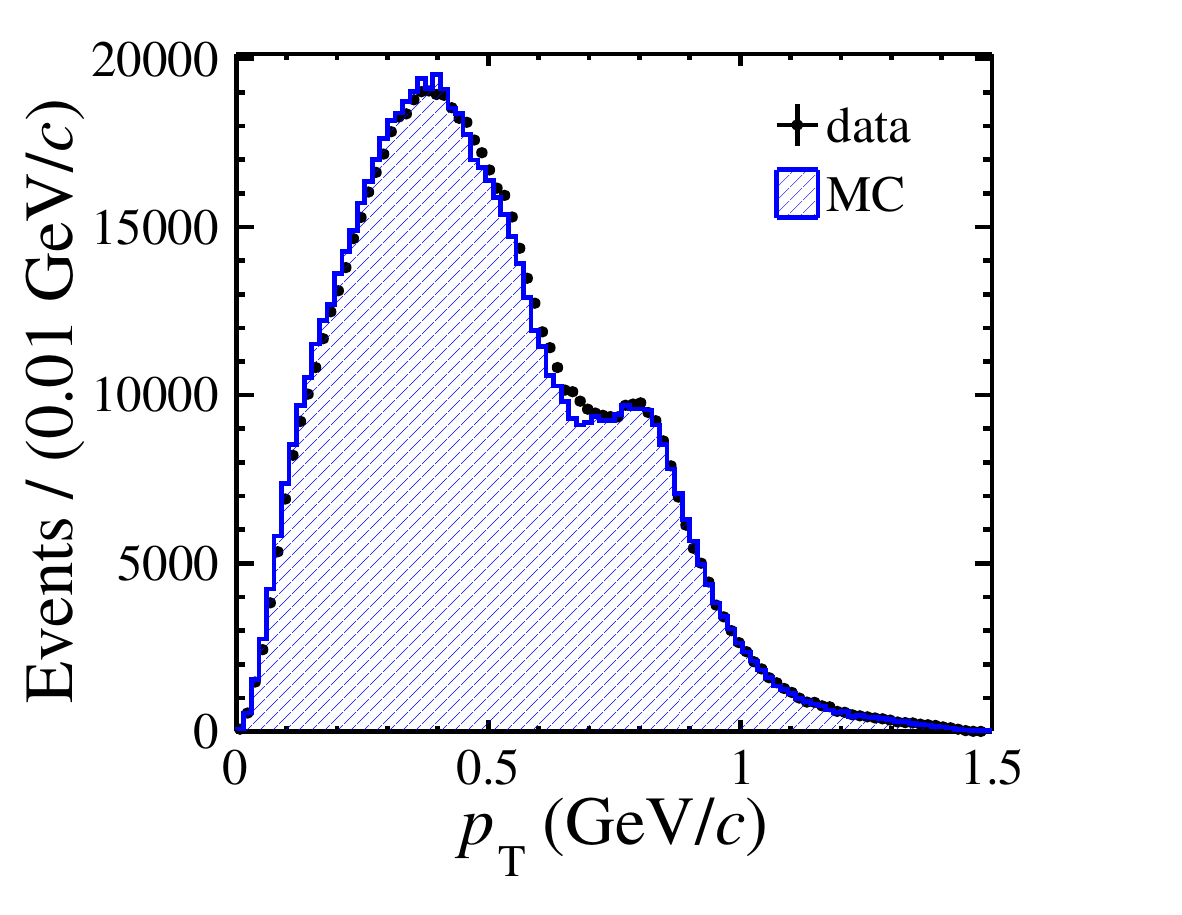}
}%
\subfigure{
\centering
\includegraphics[width=0.3\textwidth]{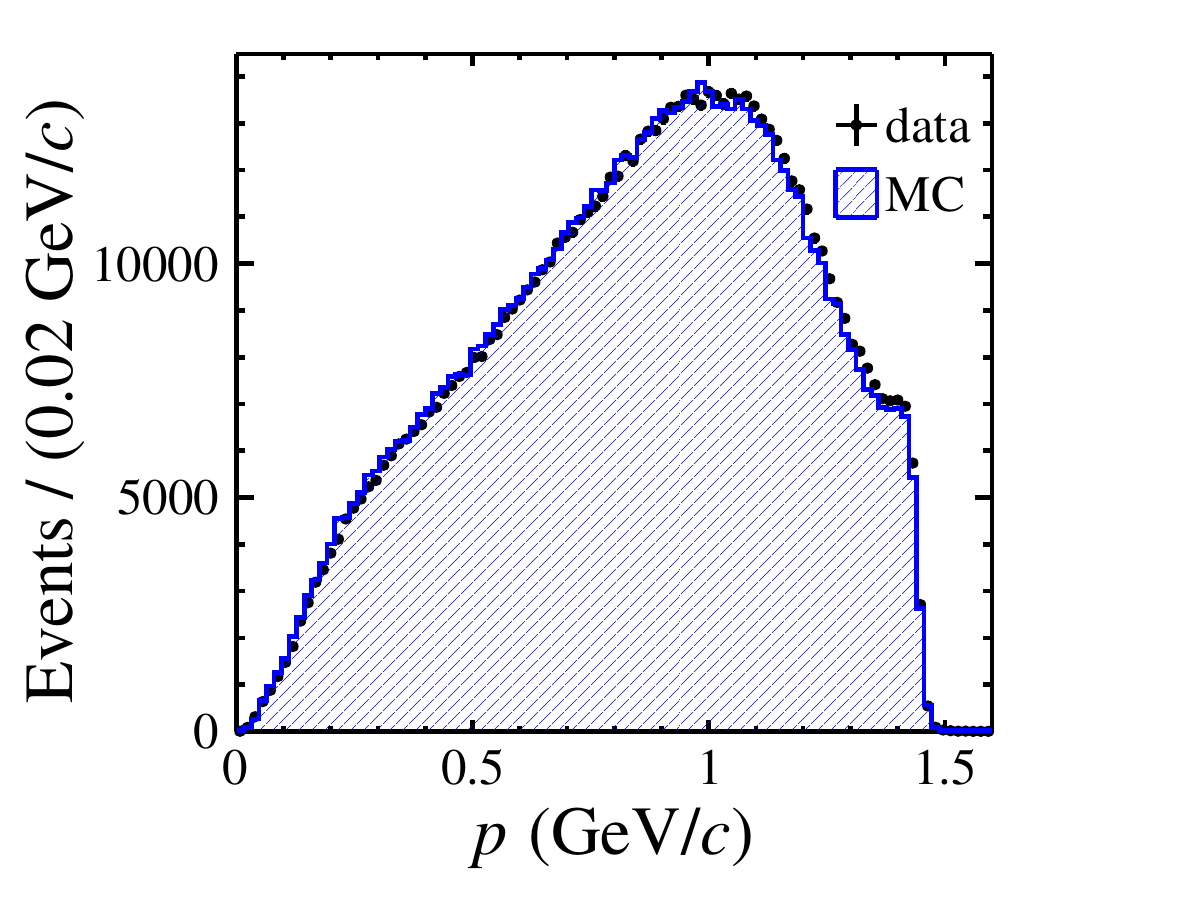}
}%
\subfigure{
\centering
\includegraphics[width=0.3\textwidth]{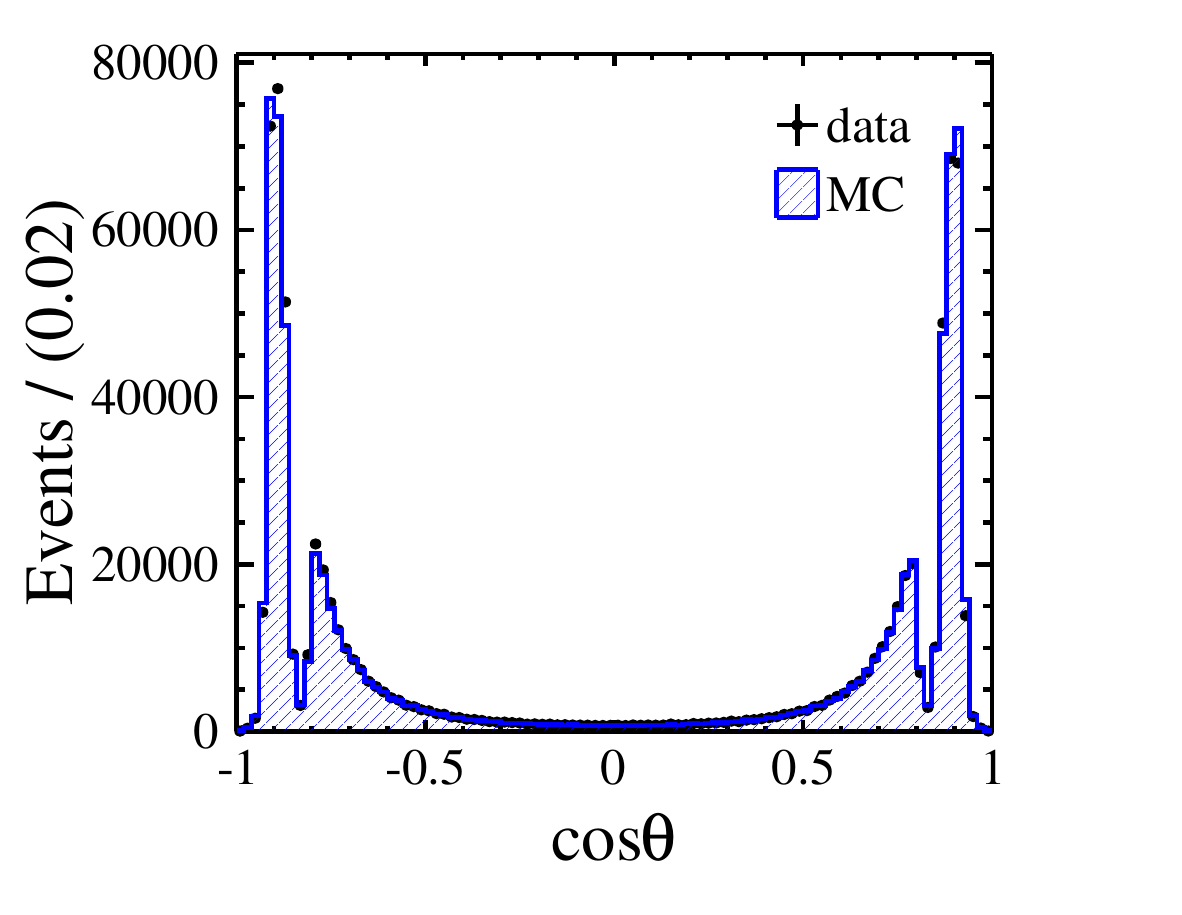}
}%
\caption{\small Distributions of transverse momentum (left),  momentum (middle) and $\cos\theta$ (right) for the probed electrons in the tracking efficiency study.}
\label{distribution_tracking}
\end{figure*}

An one-constraint kinematic fit is performed
by assuming four-momentum conservation and constraining the missing mass of the probed electron to its nominal one~\cite{pdg}. The $\chi^2$ is required to be less than 5, as shown in Fig.~\ref{evt_tracking}. The track with lower momentum is taken as the probed electron, and its four momentum is calculated assuming four-momentum conservation. In order to avoid cases where the missing track is far from the found probed electron, the matching angle between the expected missing track and the found track is required to be less than $25^\circ$. The distributions of the momentum and $\cos\theta$ of the probed electrons are shown in Fig.~\ref{distribution_tracking}.

The possible background yield is estimated by analyzing the inclusive MC sample, and the signal purity is estimated to be 99.98$\%$. The cross section of the Bhabha scattering process at $\sqrt s=3.08 \gev$ is much larger than that of other QED processes, and the MC distributions describe the data well. Therefore, the background contribution from other QED processes is negligible.

\subsection{Tracking efficiency}

The electron tracking efficiency and its systematic uncertainty are characterized in dependence of $p_T$ and $\cos\theta$, as the tracking efficiency is more sensitive to these two variables.
Although different decay processes have different systematic uncertainties for tracking, it is possible to provide a general systematic uncertainty for various decay processes according to the different momentum and angular distributions. 

\begin{figure*}[!htbp]
\centering
\subfigure[$\epsilon_{\rm data}$ of $e^-$]{
\centering
\includegraphics[width=0.5\textwidth]{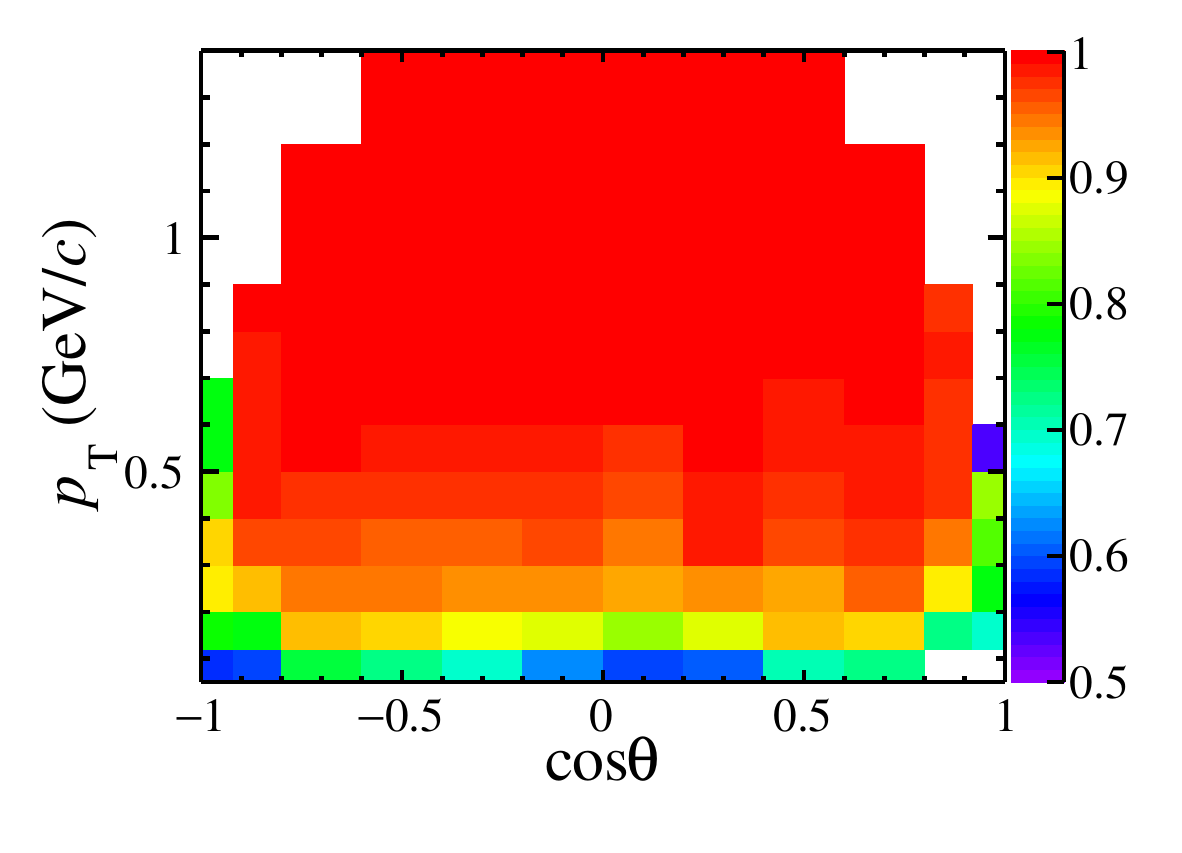}
}%
\subfigure[$\epsilon_{\rm data}$ of $e^+$]{
\centering
\includegraphics[width=0.5\textwidth]{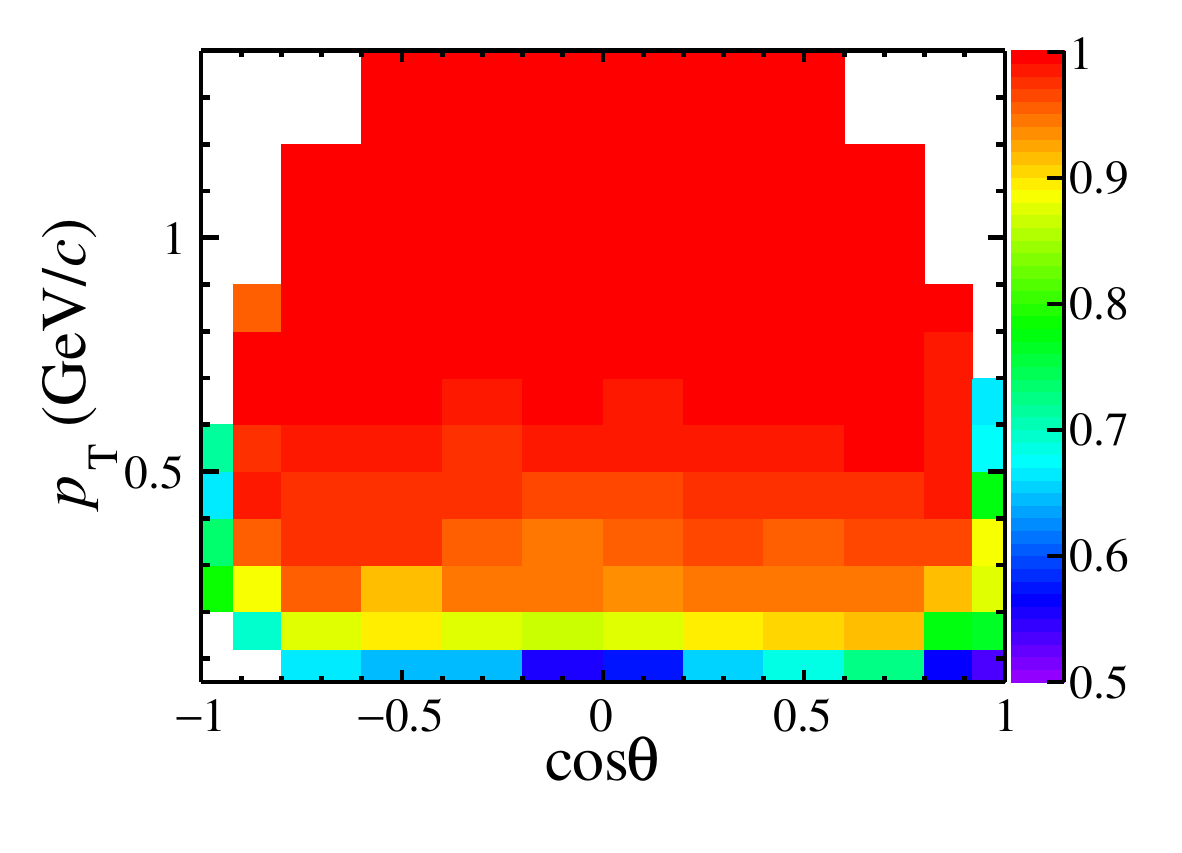}
}%

\subfigure[$\epsilon_{\rm MC}$ of $e^-$]{
\centering
\includegraphics[width=0.5\textwidth]{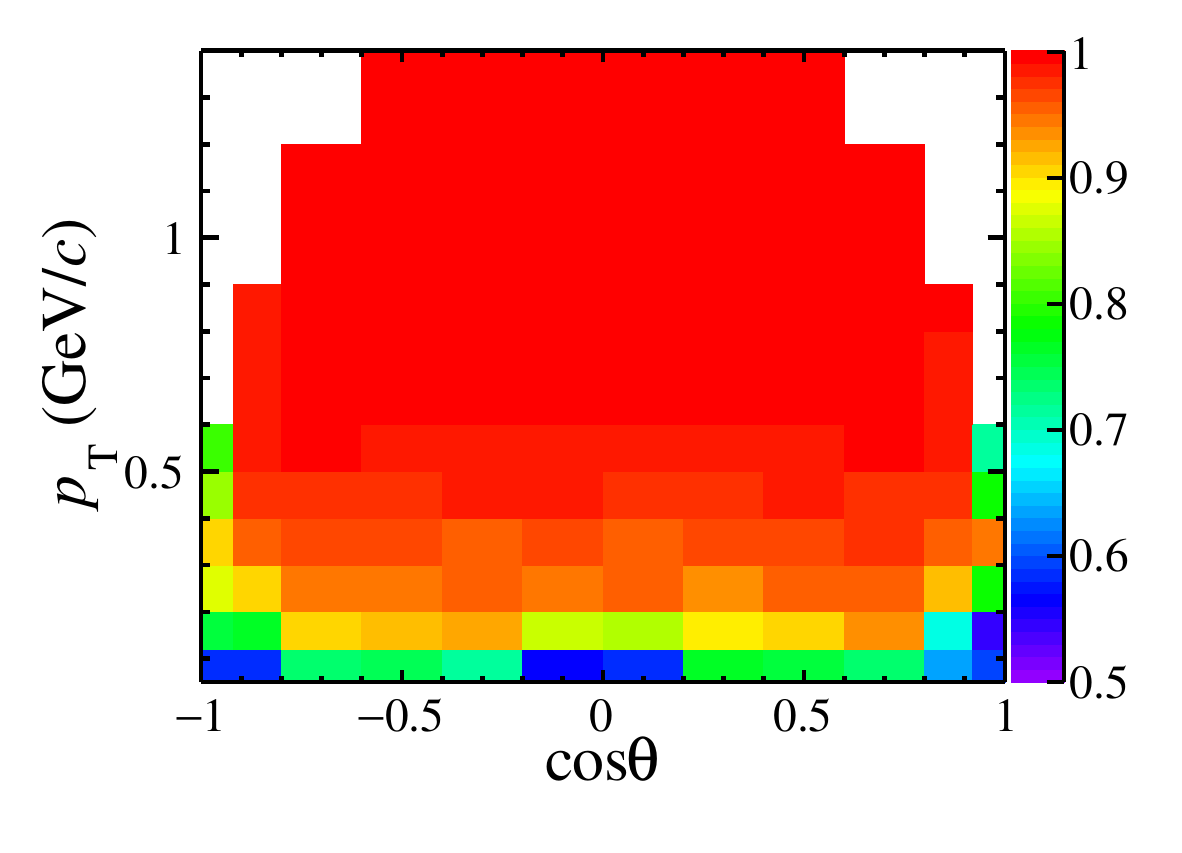}
}%
\subfigure[$\epsilon_{\rm MC}$ of $e^+$]{
\centering
\includegraphics[width=0.5\textwidth]{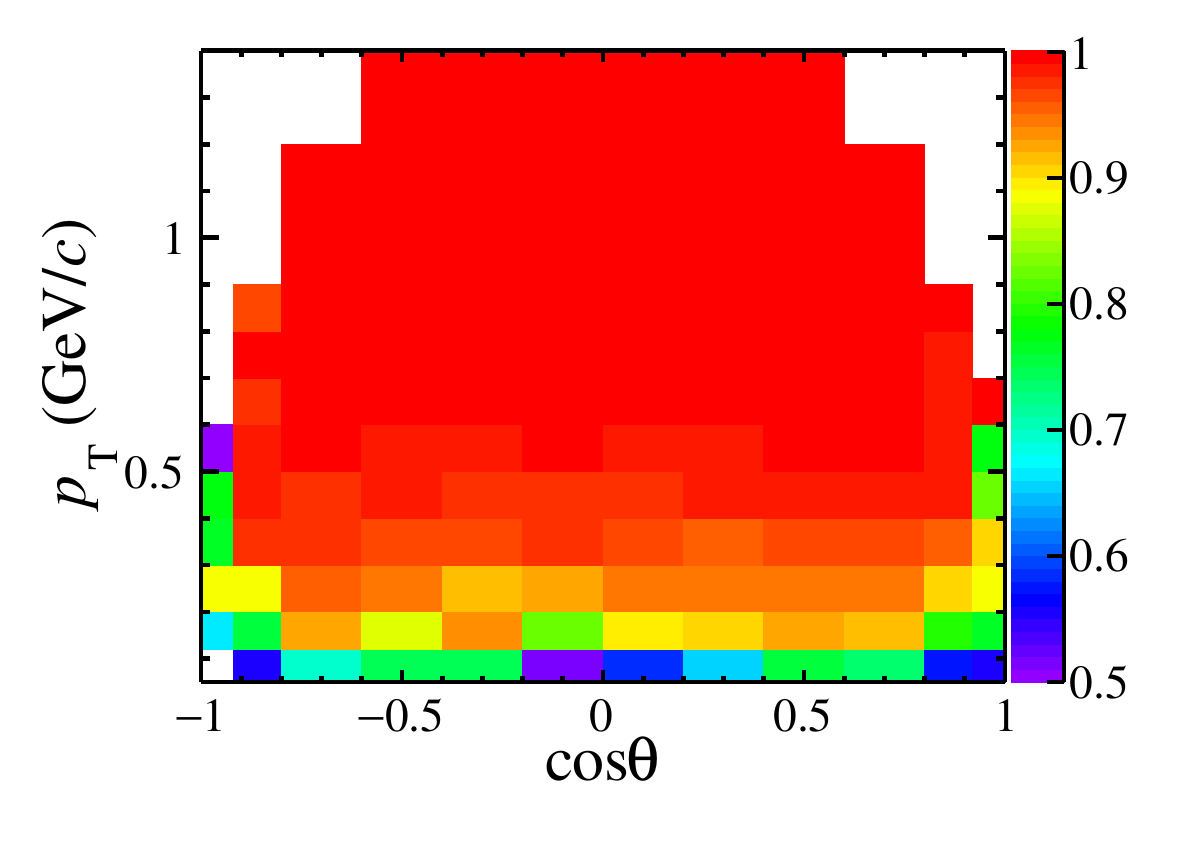}
}%

\subfigure[$\frac{\epsilon_{\rm data}}{\epsilon_{\rm MC}}$ of $e^-$]{
\centering
\includegraphics[width=0.5\textwidth]{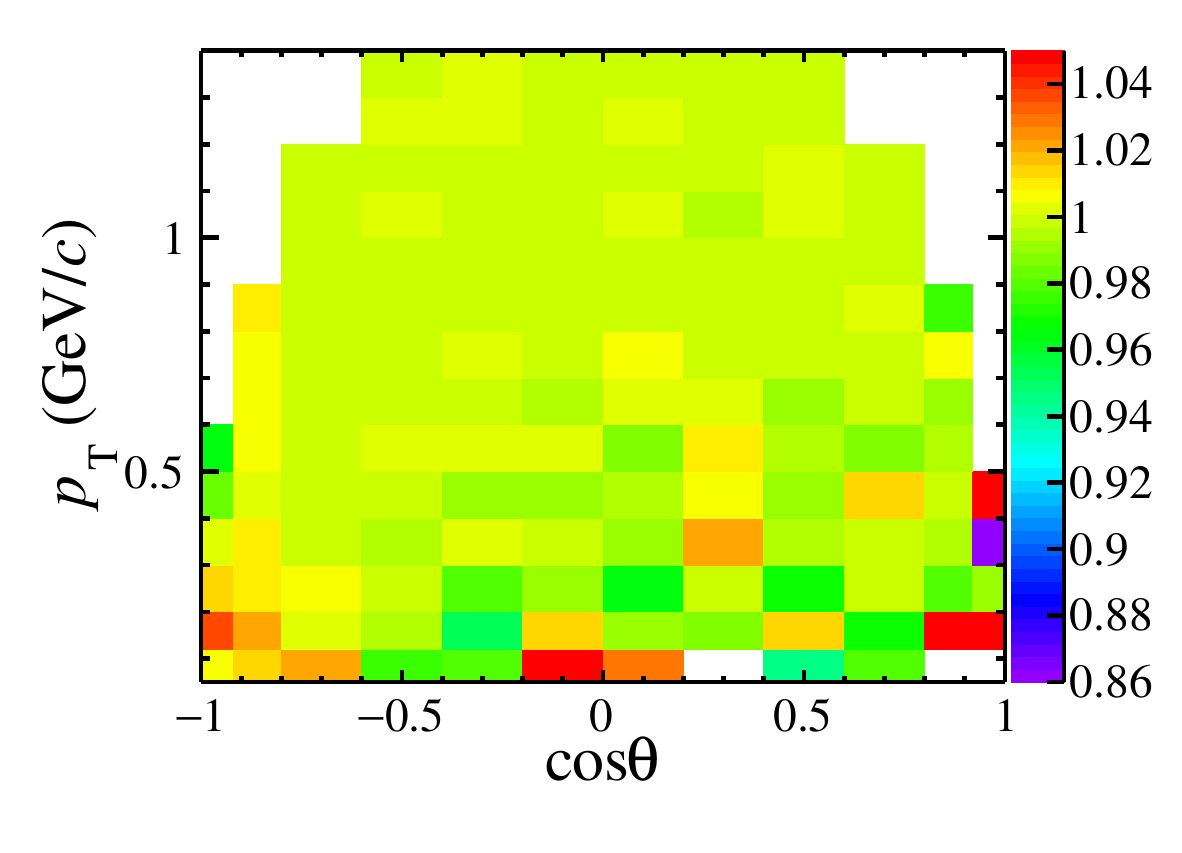}
}%
\subfigure[$\frac{\epsilon_{\rm data}}{\epsilon_{\rm MC}}$ of $e^+$]{
\centering
\includegraphics[width=0.5\textwidth]{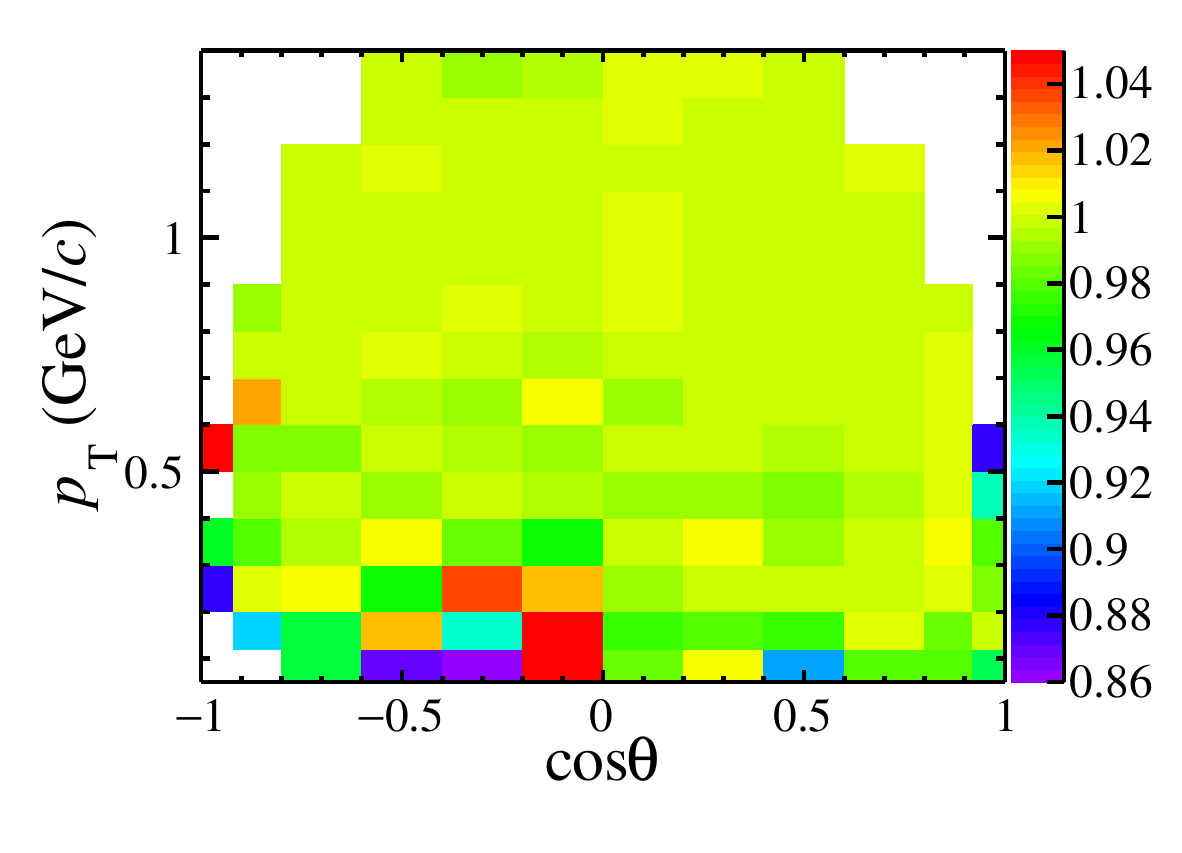}
}%
\caption{\small Electron (left) and positron (right) tracking efficiencies of data (top) and  MC (middle) as well as the correction factor (bottom) versus $p_T$ and $\cos\theta$. }
\label{eff_tracking}
\end{figure*}

\begin{figure*}[!htbp]
\centering
\subfigure[$\alpha$ of $e^-$]{
\centering
\includegraphics[width=0.5\textwidth]{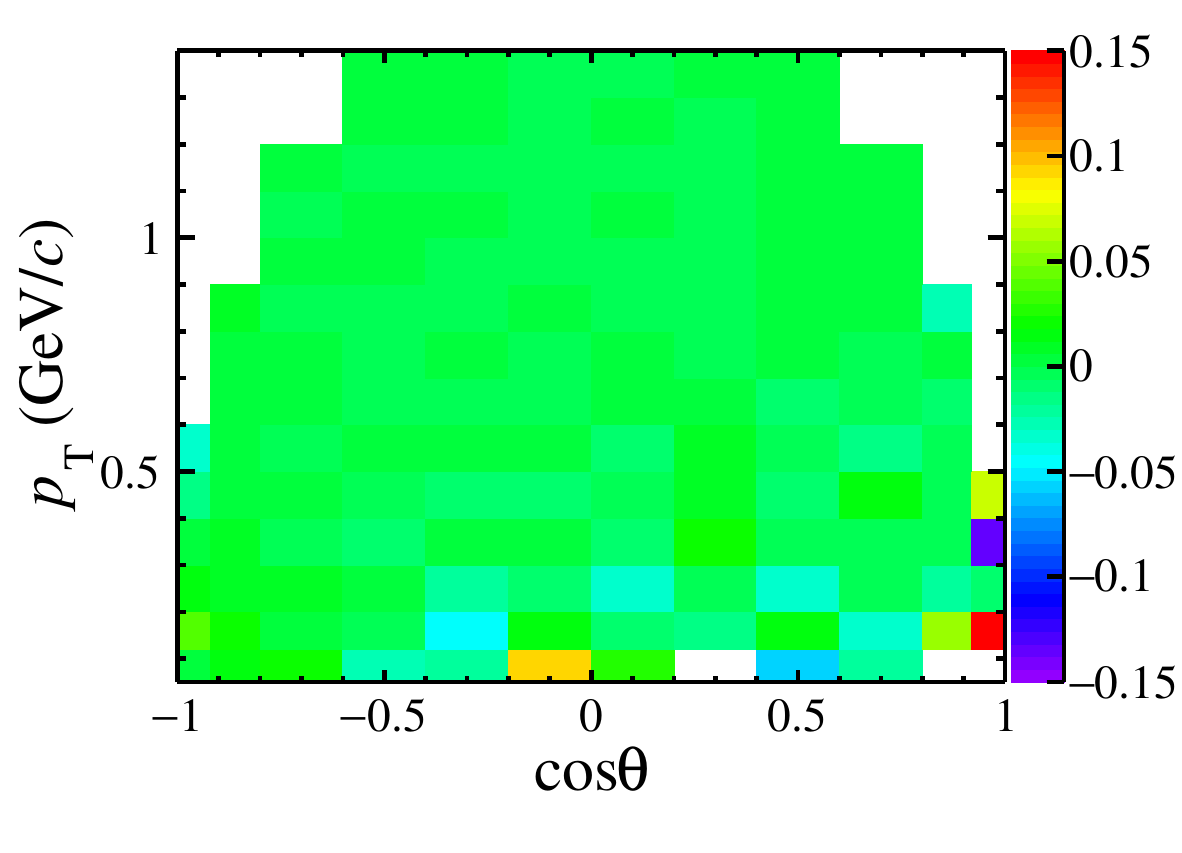}
}%
\subfigure[$\sigma_\alpha$ of $e^-$]{
\centering
\includegraphics[width=0.5\textwidth]{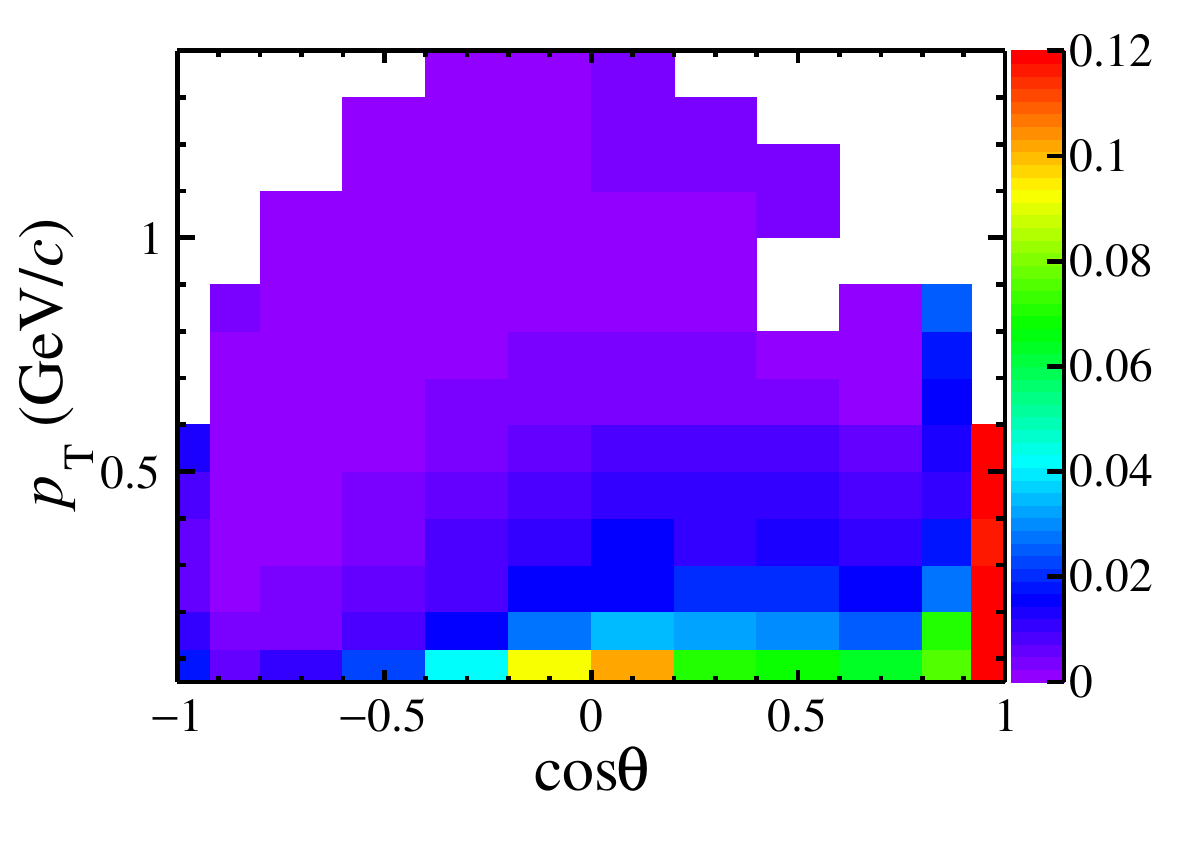}
}%

\subfigure[$\alpha$ of $e^+$]{
\centering
\includegraphics[width=0.5\textwidth]{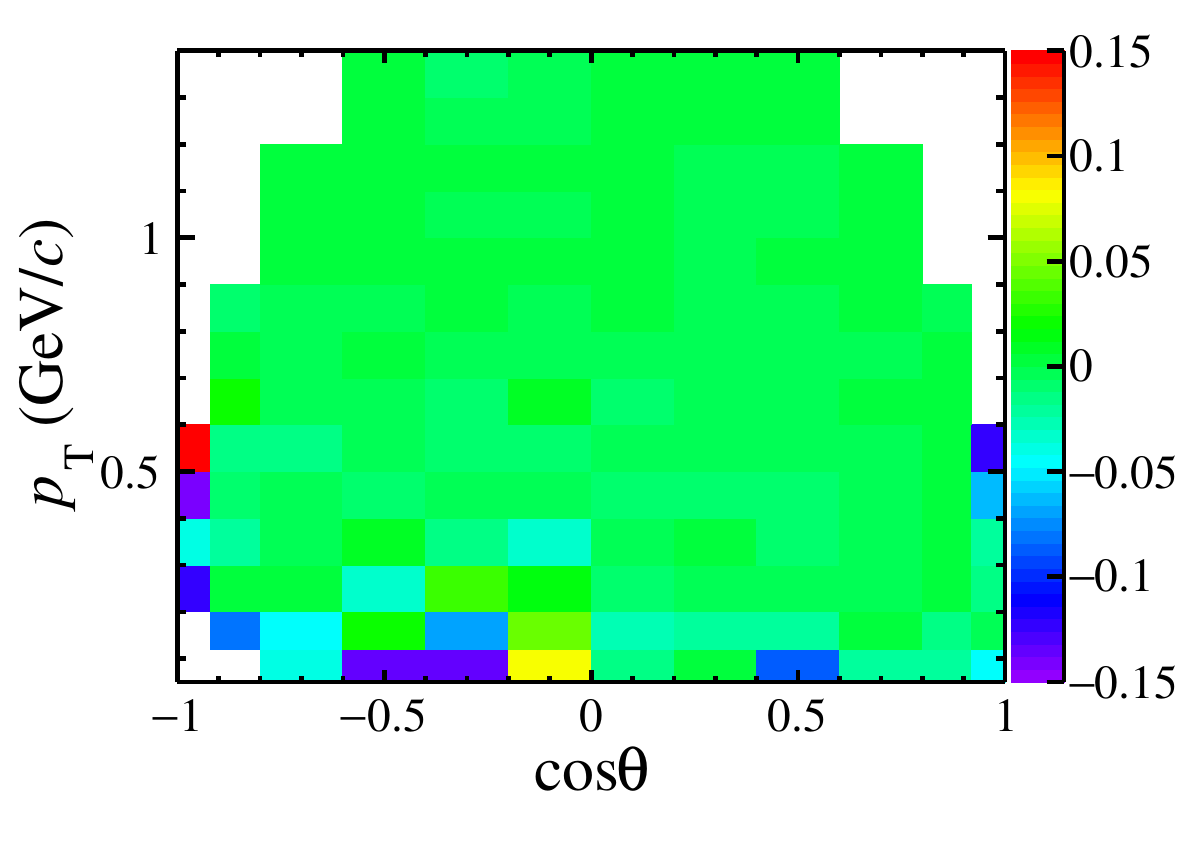}
}%
\subfigure[$\sigma_\alpha$ of $e^+$]{
\centering
\includegraphics[width=0.5\textwidth]{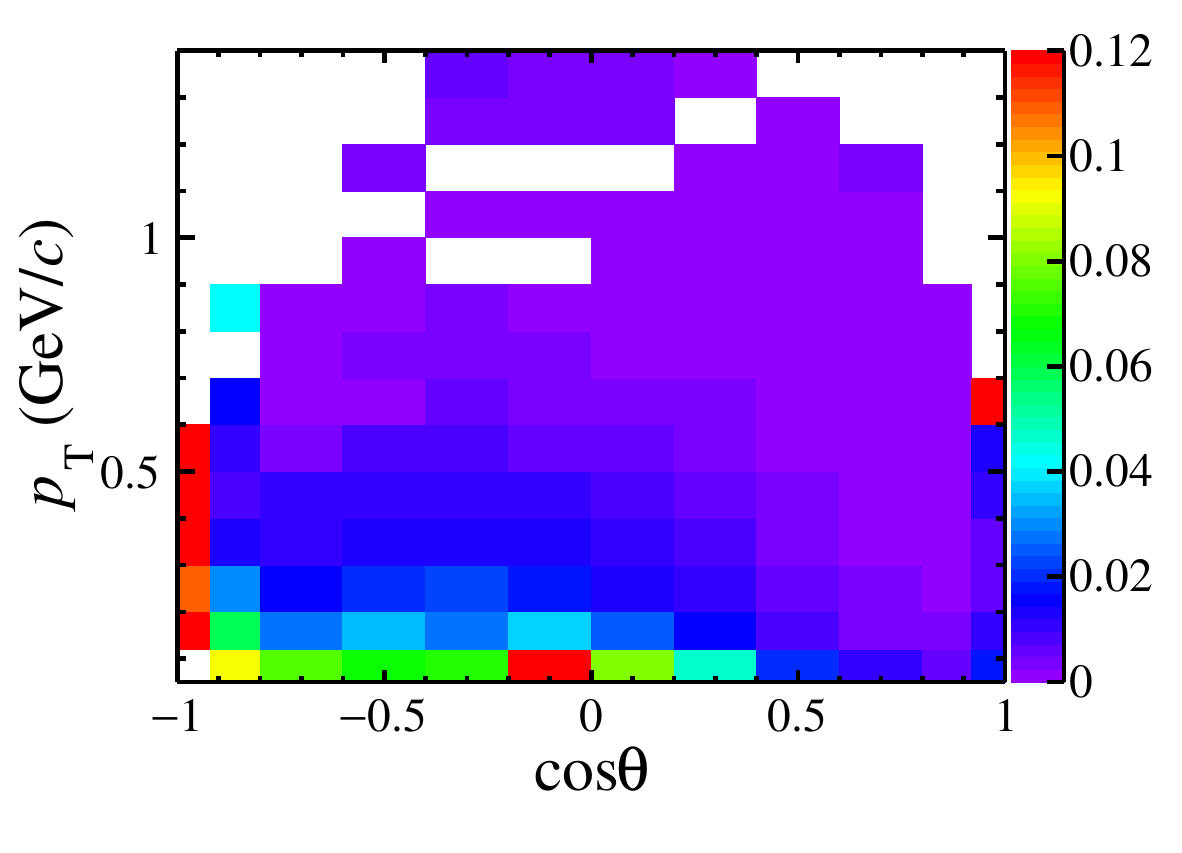}
}%
\caption{\small Relative difference in the electron (top row) and positron (bottom row) tracking efficiencies  between data and MC (left) and the uncertainty of the correction factor (right) versus $p_{T}$ and $\cos\theta$.}
\label{2d_sys_tracking_18_19}
\end{figure*}

The efficiencies of data and MC as well as correction factors are demonstrated in Fig.~\ref{eff_tracking}. The relative difference of $\epsilon_{\rm data}$ and $\epsilon_{\rm MC}$ is represented in two dimensions, $p_T$ versus $\cos\theta$. 
An example of such a two-dimensional distribution is illustrated in Fig.~\ref{2d_sys_tracking_18_19} for electrons, as a function of $p_T$ and $\cos\theta$, where the colors in each box represent the corresponding relative difference of tracking efficiency between data and MC.

The relative differences in electron tracking efficiencies between data and MC are less than 0.5$\%$ in most bins.  The correction factors for electrons and positrons are different in most of low momentum bins, while they are both close to 100\% in high momentum bins.
If 
an analysis further applies the correction factors in the tracking efficiency correction, the systematic uncertainty due to tracking efficiency will be even smaller.
This is important for high precision measurements at the BESIII experiment~\cite{Asner:2008nq, BESIII:2020nme}.

\section{PID efficiency study}

\subsection{Control sample selection}\label{sec::pid}

The candidate events for $\ee\ar\ee\gamma$ are required to have two good charged tracks with a total charge of zero. The requirements on the polar angle and the point of closest approach to the interaction point are the same as those used in the tracking efficiency study. The track with momentum greater than 1.47 GeV, corresponding to at least 95$\%$ of the beam momentum, is considered the `tagged track'. 
For the tagged track, the ratio $E/p$ is required to be in the range from 0.8 to 1.2. 
Additionally, a vertex fit is performed to ensure that electron and positron originate from the same vertex.

\begin{figure}[!htbp]
\centering
\includegraphics[width=0.4\textwidth]{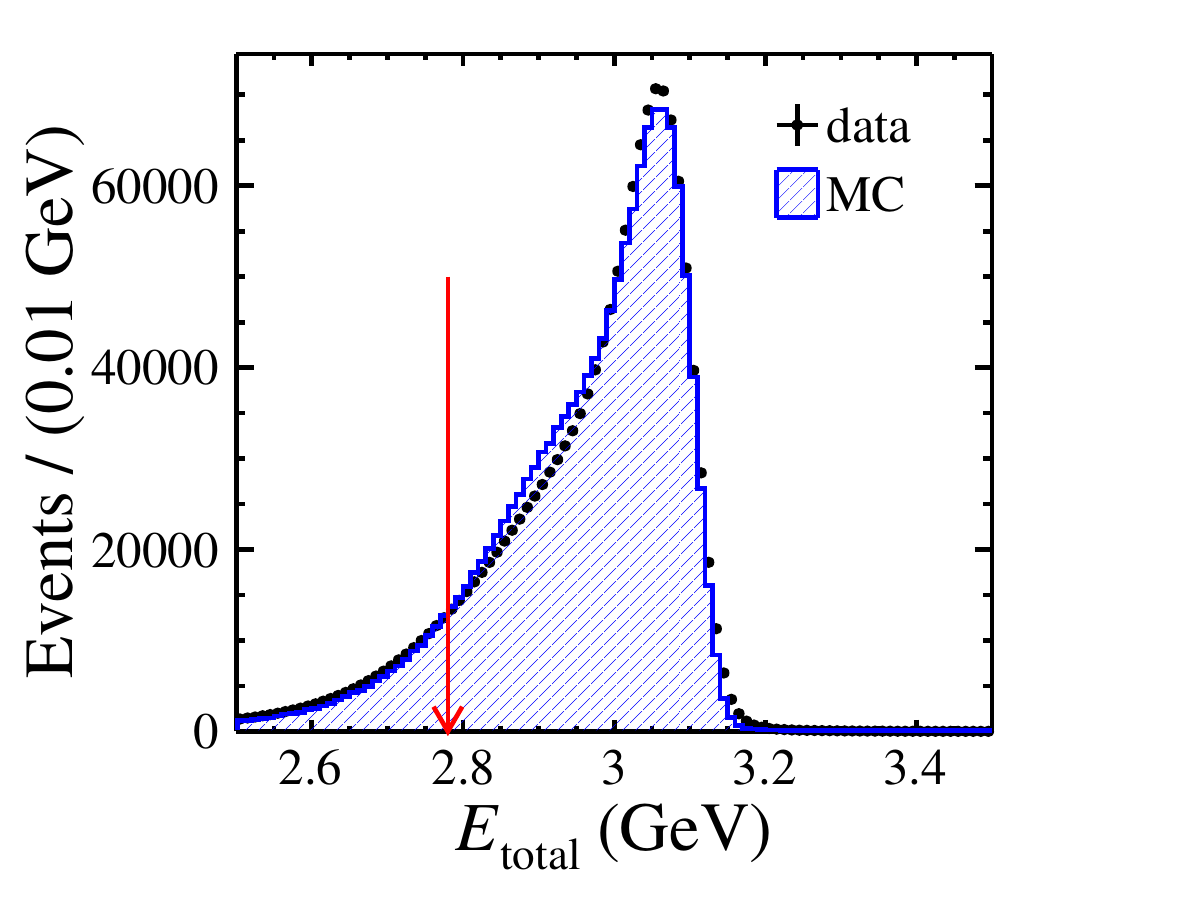}
\caption{\small Distribution of the total energy deposited in the EMC for selected data and MC samples in PID efficiency study.}
\label{evt_PID}
\end{figure}

\begin{figure*}[!htbp]
\centering
\subfigure{
\centering
\includegraphics[width=0.3\textwidth]{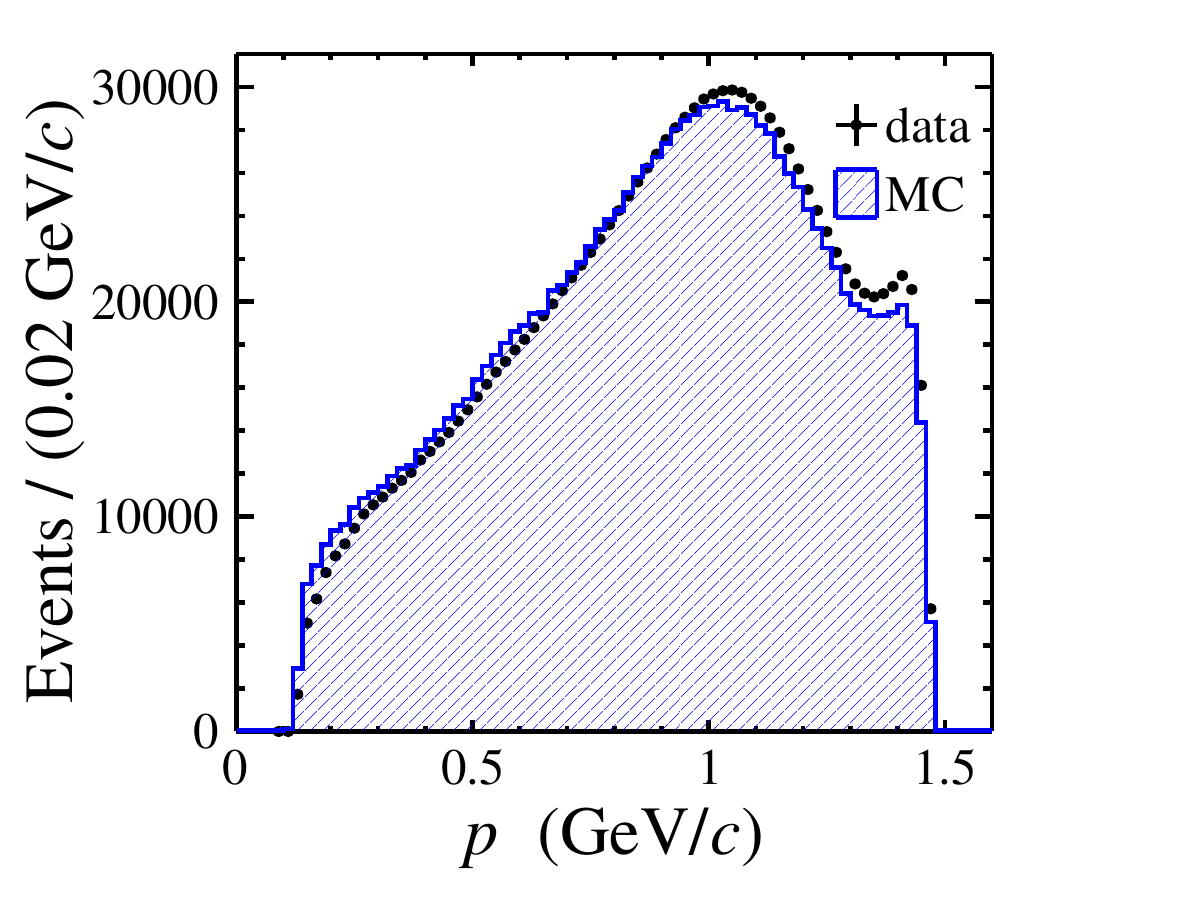}
}%
\subfigure{
\centering
\includegraphics[width=0.3\textwidth]{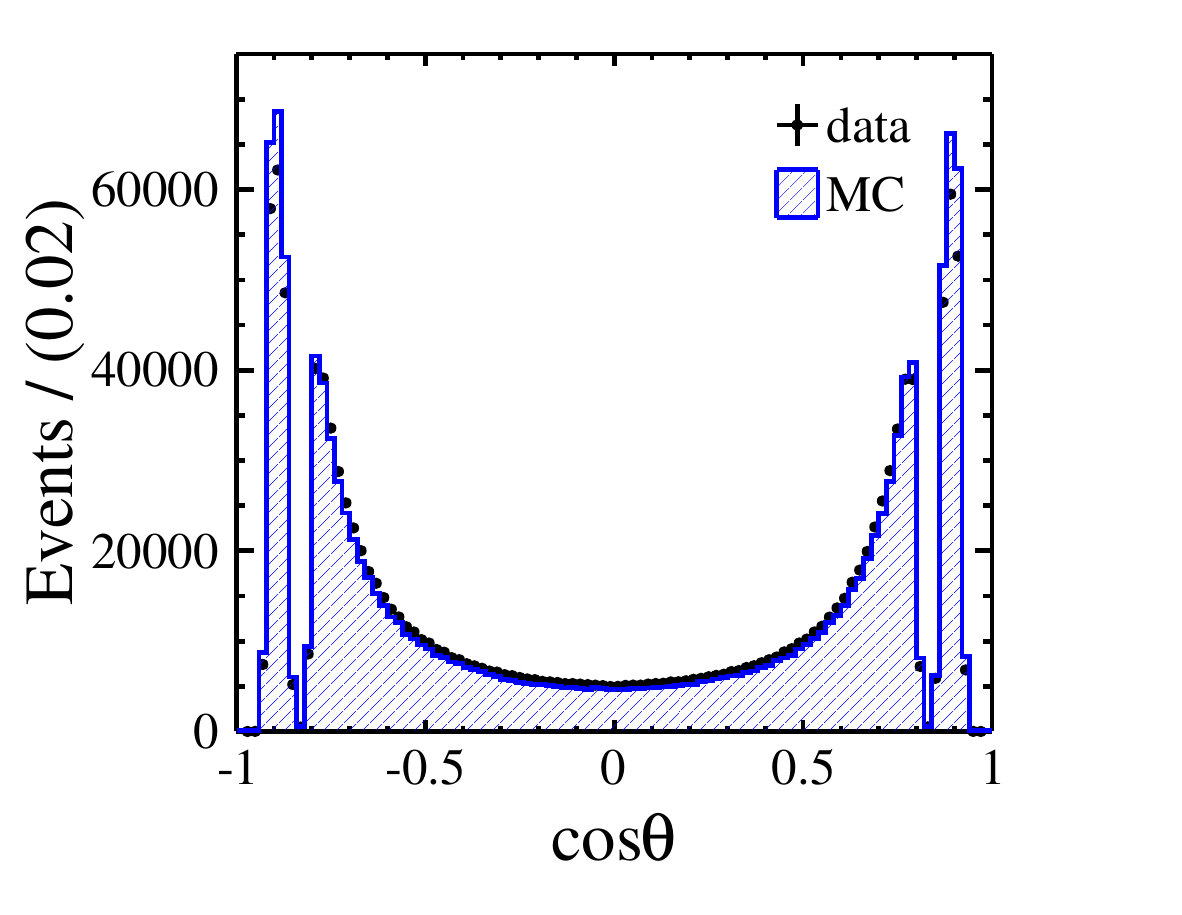}
}%
\subfigure{
\centering
\includegraphics[width=0.3\textwidth]{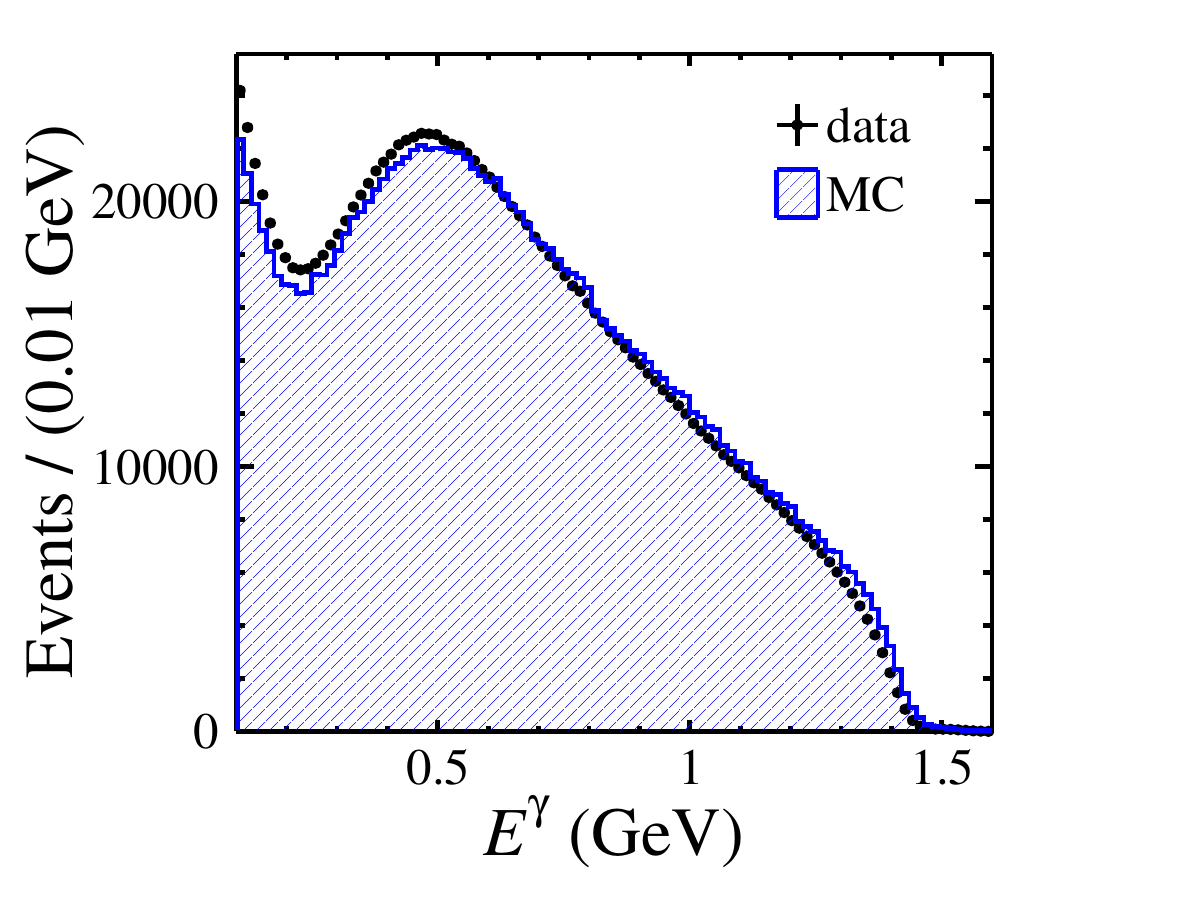}
}%
\caption{\small Distributions of momentum (left) and $\cos\theta$ (middle) for the probed electrons and photon energy (right) in PID efficiency study.}
\label{distribution_PID}
\end{figure*}

The event selection criteria for photon candidates are the same with those used in the tracking efficiency study. The total energy deposit in the EMC is required to be greater than 2.78 GeV, corresponding to 90$\%$ of the center-of-mass energy, as shown in Fig.~\ref{evt_PID}. Good consistency between data and MC can be seen. The distributions of the momentum and $\cos\theta$ from the probed electrons are illustrated in Fig.~\ref{distribution_PID}.

The possible background events are estimated through analyzing the inclusive MC sample.  This study shows a signal purity of 99.99$\%$.

\subsection{PID efficiency}

Electron PID uses the measured information in the MDC, TOF and EMC. The combined likelihoods under the electron, pion, kaon, and proton hypotheses are obtained.
The PID efficiency and its systematic uncertainty are characterized dependent on the track's momentum $p$ and its polar angle $\cos\theta$.

Considering different momentum and angular distributions in different physics analyses, we have studied two alternative options to derive the confidence levels: 
A) calculating the PID information by combining the information from d$E$/d$x$ and TOF, mainly for low-momentum electrons that may not reach the EMC.
B) utilizing the full PID information from d$E$/d$x$, TOF and EMC.  In both cases, tracks are considered as electron if they satisfy $\frac{CL(e)}{CL(e)+CL(\pi)+CL(K)}>0.8$; this requirement is commonly used to suppress potential $\pi$ and $K$ backgrounds. 

The option A) has poor performance in the region of small $|\cos\theta|$ due to the short flight path and limited ionization sampling of the charged track.  Therefore, incorporating information from the EMC can significantly enhance its performance. 
Option B) uses more information and has better performance in suppressing wrong PID assignments.

The choice of the PID strategy depends on the specific analysis. If the electrons can reach the EMC, it is highly recommended to use the information from the EMC which significantly improves the PID efficiency  for electrons with small $|\cos\theta|$. On the other hand, the option B) is suggested when mis-identification contributes predominantly to background and may have better performance in such cases. Furthermore, a requirement for the absolute value of $P(e)$ could also be considered in the option B).

\begin{figure*}[!htbp]
\centering
\subfigure[$\epsilon_{\rm data}$ of $e^-$]{
\centering
\includegraphics[width=0.5\textwidth]{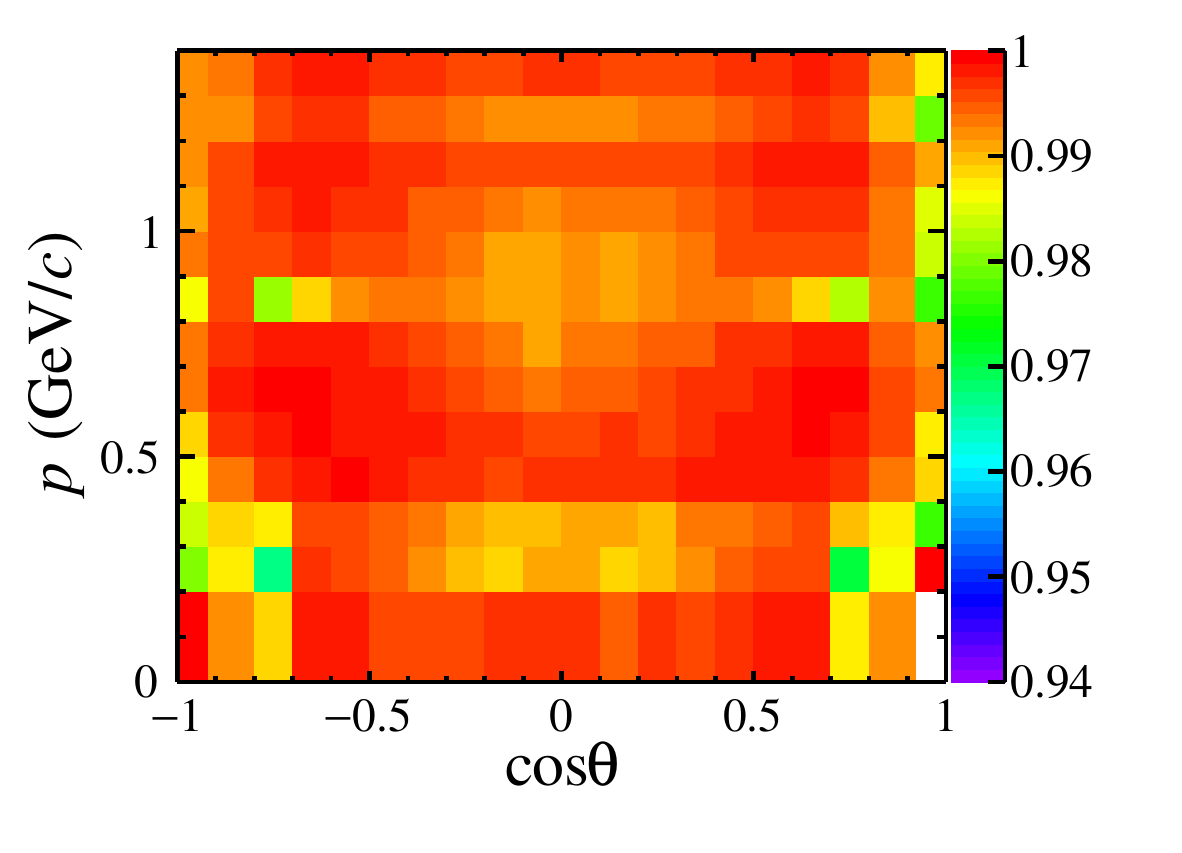}
}%
\subfigure[$\epsilon_{\rm data}$ of $e^+$]{
\centering
\includegraphics[width=0.5\textwidth]{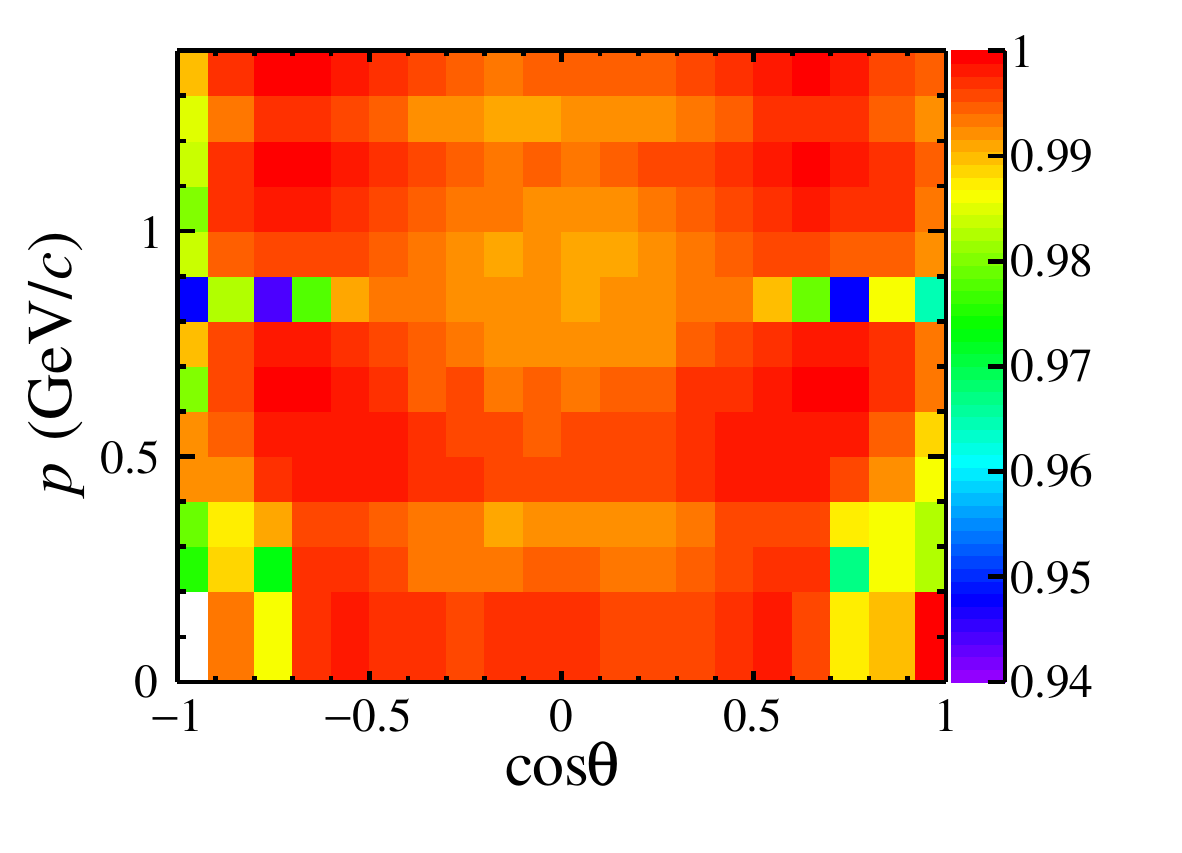}
}%

\subfigure[$\epsilon_{\rm MC}$ of $e^-$]{
\centering
\includegraphics[width=0.5\textwidth]{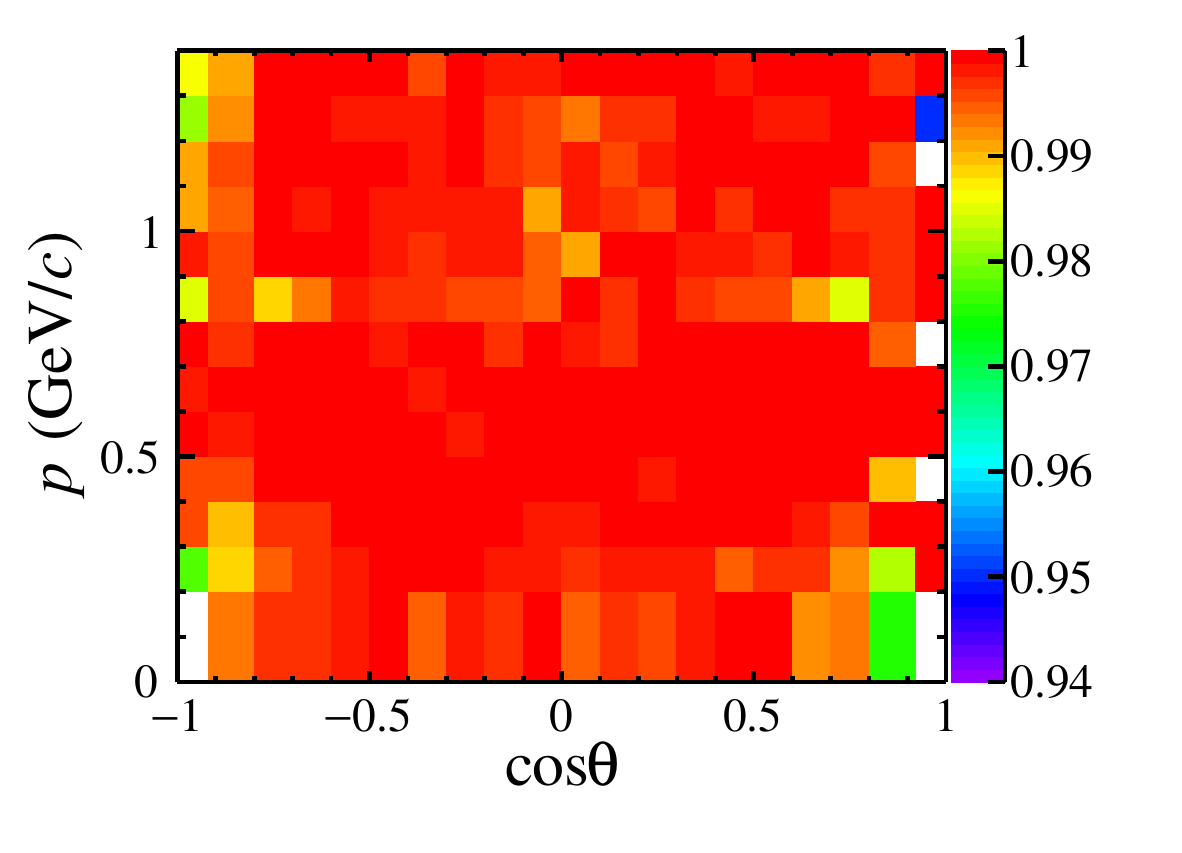}
}%
\subfigure[$\epsilon_{\rm MC}$ of $e^+$]{
\centering
\includegraphics[width=0.5\textwidth]{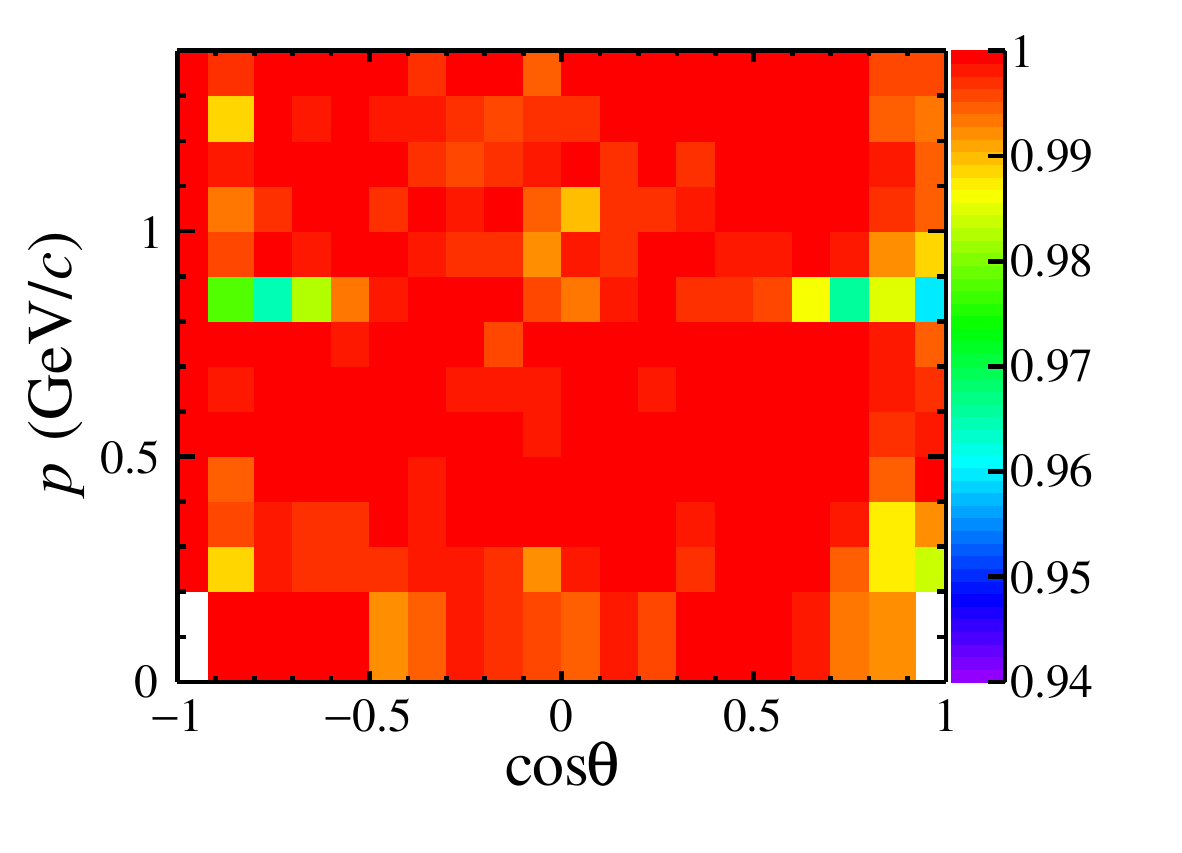}
}%

\subfigure[$\frac{\epsilon_{\rm data}}{\epsilon_{\rm MC}}$ of $e^-$]{
\centering
\includegraphics[width=0.5\textwidth]{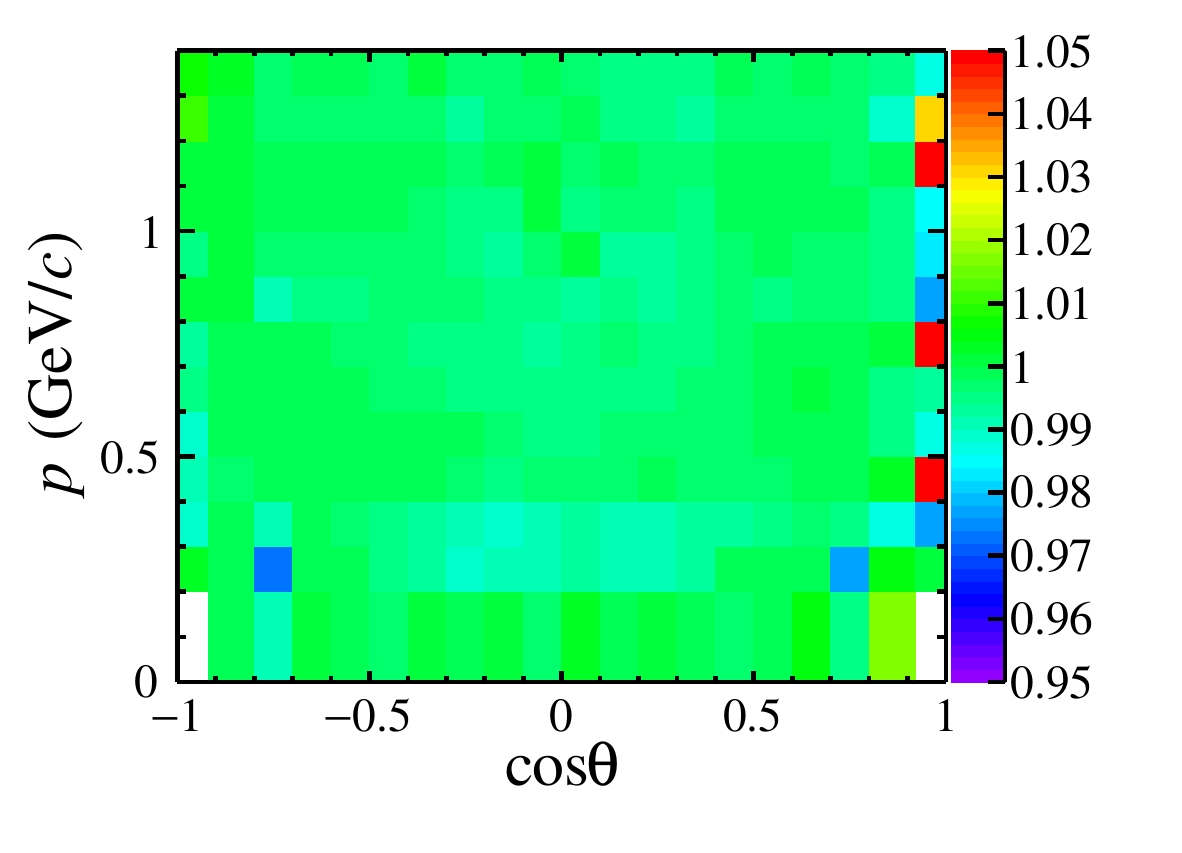}
}%
\subfigure[$\frac{\epsilon_{\rm data}}{\epsilon_{\rm MC}}$ of $e^+$]{
\centering
\includegraphics[width=0.5\textwidth]{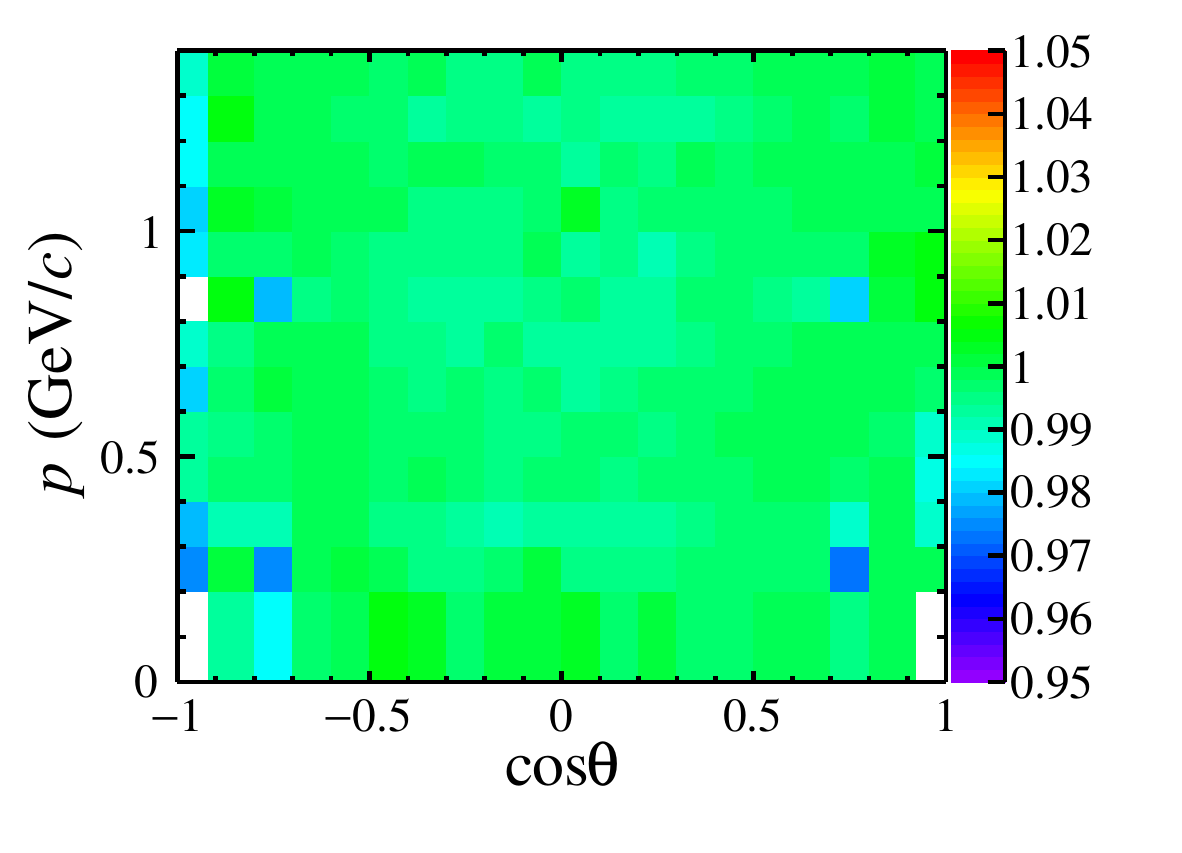}
}%
\caption{\small Electron (left) and positron (right) PID efficiencies of data (top) and MC (middle) as well as the correction factor (bottom) versus $p$ and $\cos\theta$.  }
\label{eff_PID}
\end{figure*}

\begin{figure*}[!htbp]
\centering
\subfigure[$\alpha$ of $e^-$ ]{
\centering
\includegraphics[width=0.5\textwidth]{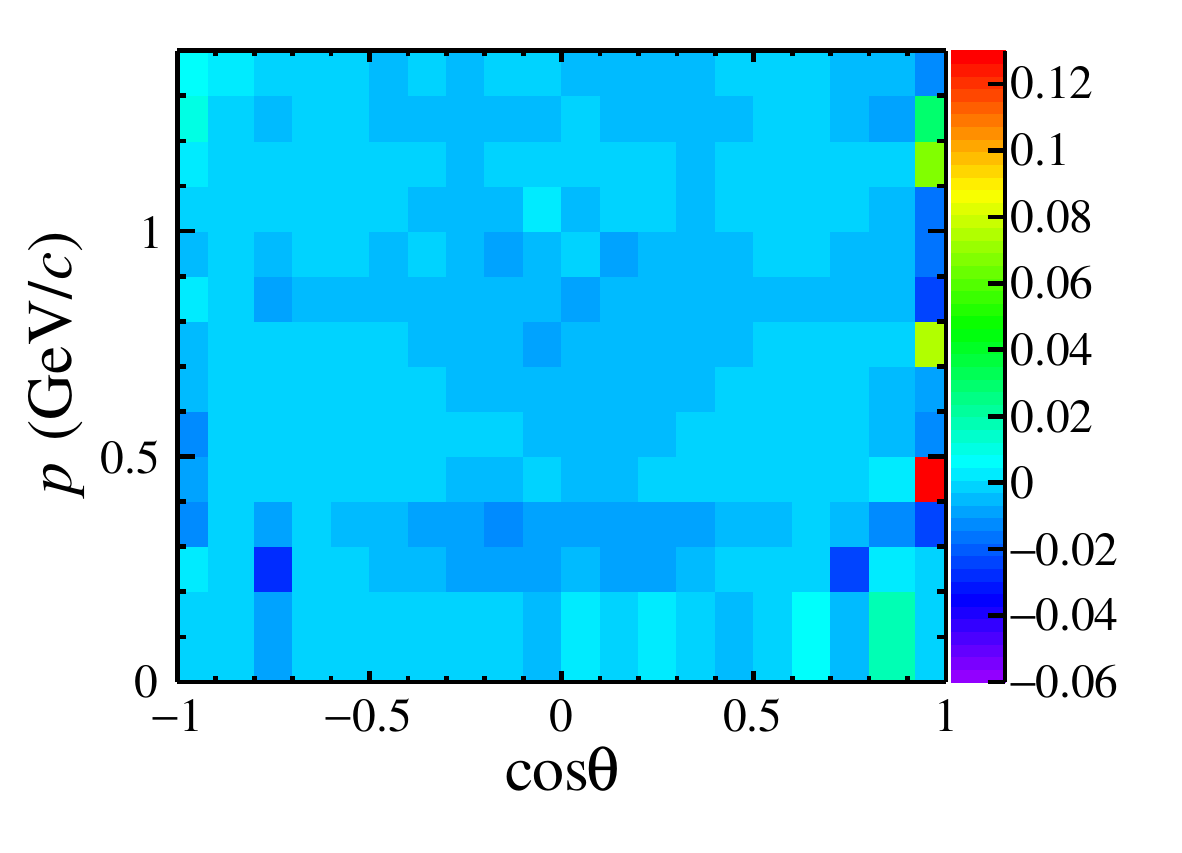}
}%
\subfigure[$\sigma_\alpha$ of $e^-$]{
\centering
\includegraphics[width=0.5\textwidth]{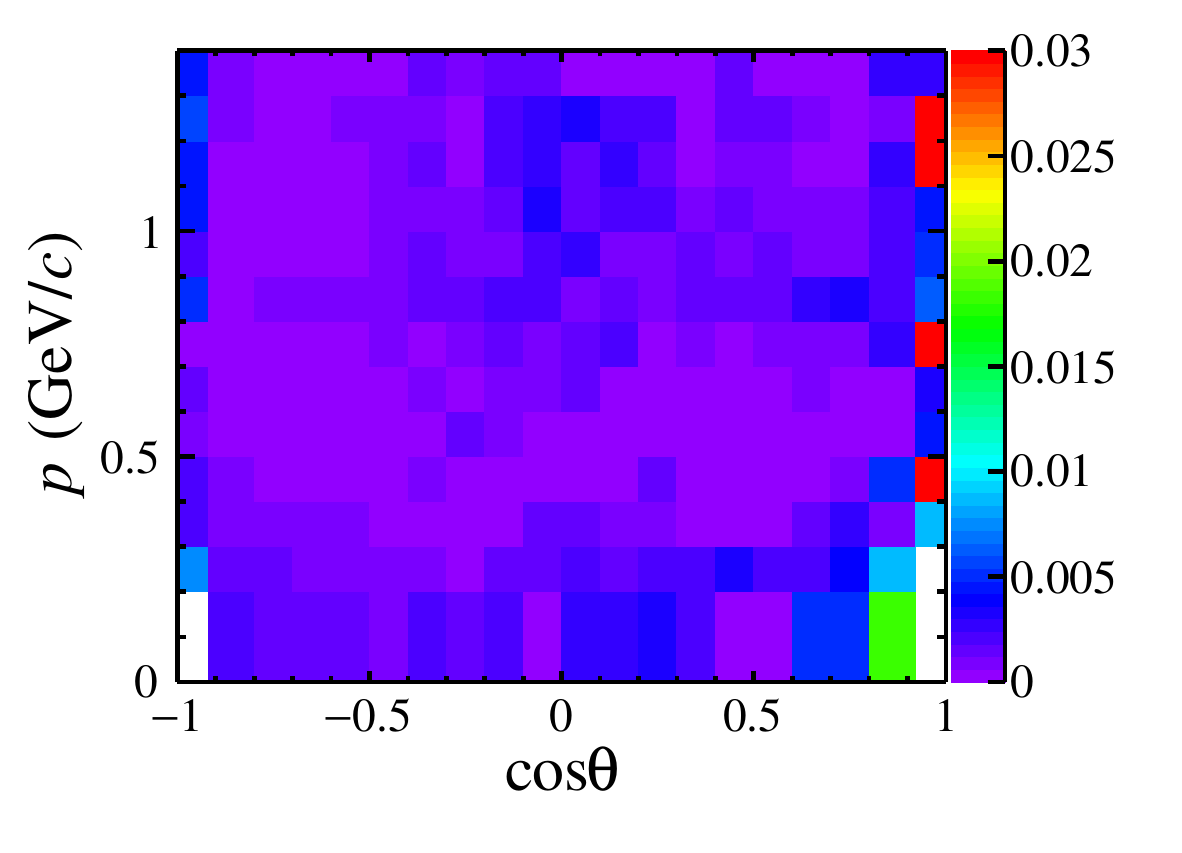}
}%

\subfigure[$\alpha$ of $e^+$ ]{
\centering
\includegraphics[width=0.5\textwidth]{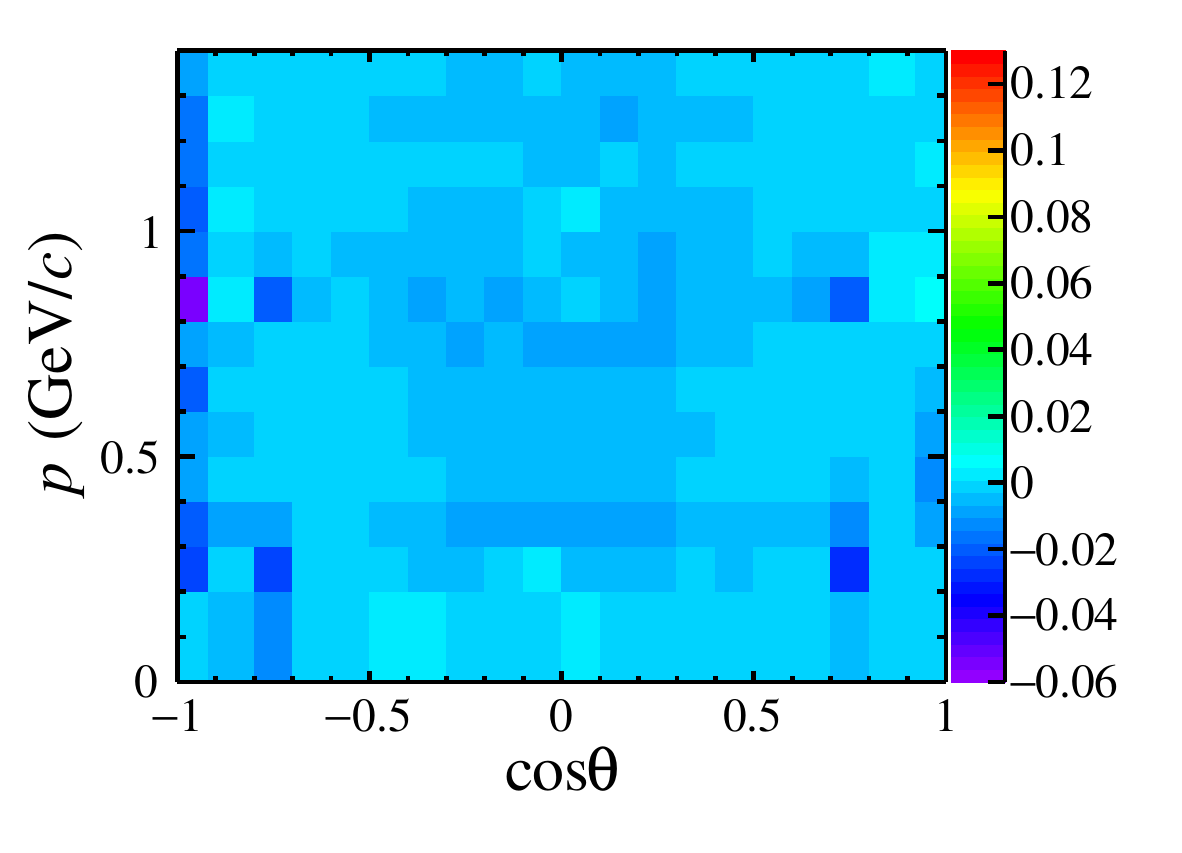}
}%
\subfigure[$\sigma_\alpha$ of $e^+$]{
\centering
\includegraphics[width=0.5\textwidth]{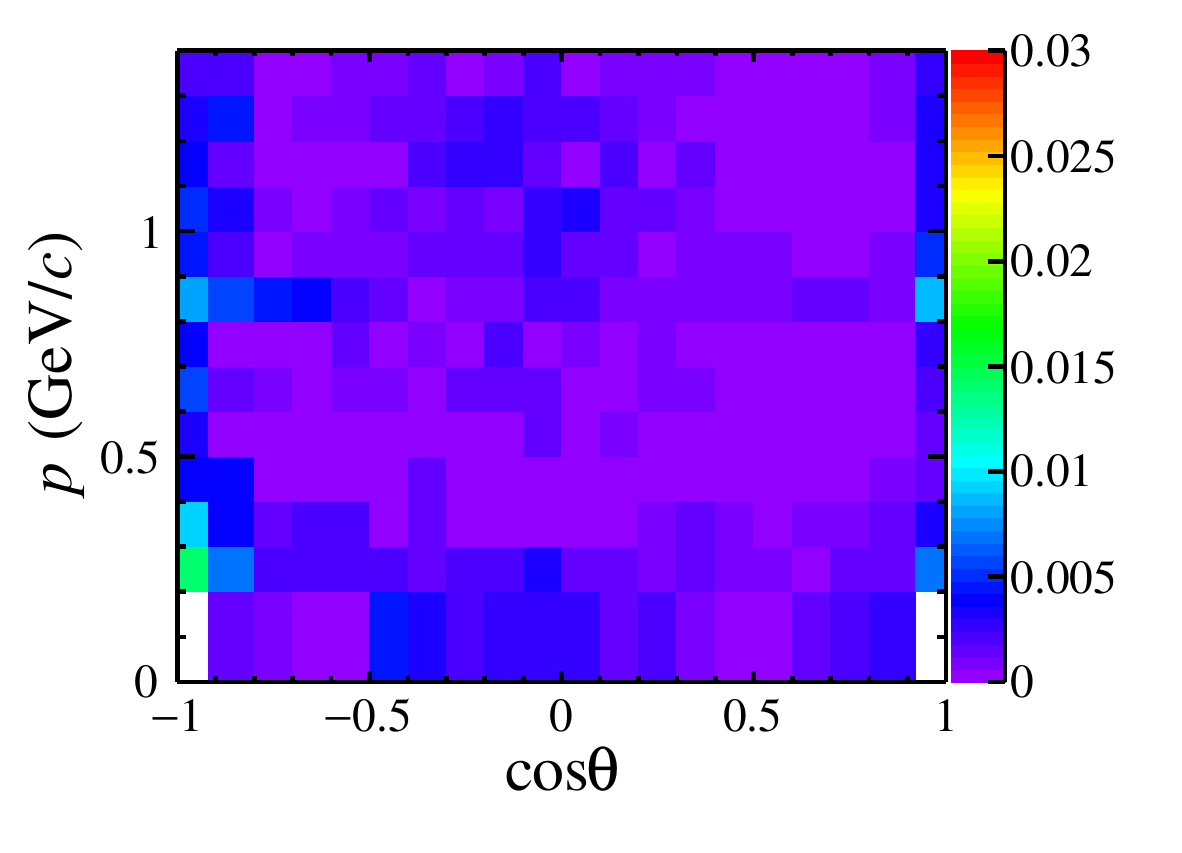}
}%
\caption{\small Relative difference in the electron (top row) and positron (botton row) PID efficiencies between data and MC (left) and the uncertainty of correction factor (right) versus $p$ and $\cos\theta$.  }
\label{reff_PID}
\end{figure*}

The efficiencies of data and MC, using the full information as in option B), as well as their ratio are shown in Fig.~\ref{eff_PID}. Additionally, the relative difference between data and MC depending on $p$ and $\cos\theta$ has been extracted, as shown in Fig.~\ref{reff_PID}. The derived correction factor can also be used in the PID efficiency correction.

The differences in the PID efficiencies between data and MC are less than 0.5$\%$ in most bins, while they are worse in low momentum bins.  The correction factors for electrons and positrons are nearly the same in most bins.
After application of the correction factors, simulated data will match real data better, and the systematic uncertainties due to differences in PID efficiencies will be less than 0.1$\%$ in most bins.

\section{Summary}
Using a data sample with an integrated luminosity of (136.22 $\pm$ 0.09) pb$^{-1}$ collected at $\sqrt s=3.08 \gev$ and a data sample with 
an integrated luminosity of (2568.07 $\pm$ 0.40) pb$^{-1}$ 
collected at $\sqrt s=3.097 \gev$, we study the electron tracking and PID efficiencies, respectively.
The systematic uncertainties of electron tracking or PID and the corresponding correction factors versus (transverse) momentum and polar angle are evaluated, indicating that the systematic uncertainties after correction are  mostly less than 0.5$\%$ within $p_T>0.4$ GeV and all momentum region, respectively.
Correction factors for tracking and PID efficiencies versus $p_T/p$ and $\cos\theta$ are been presented that can be used in data analysis of other processes to further reduce systematic uncertainties.

\section{Acknowledgments}
This work is partially supported by National Key R$\&$D Program of China under Contracts No. 2020YFA0406404 and Joint Large-Scale Scientific Facility Funds of the NSFC and CAS under Contract No. U1832207.

\end{document}